# Rock-bridge mechanics: A review

Fengchang Bu[1,2,3*], Michel Jaboyedoff[3], Lei Xue[4]

1. Department of Earth Science and Engineering, Imperial College London, London, UK
2. Department of Civil and Environmental Engineering, Massachusetts Institute of Technology, Cambridge, MA, USA
3. Institute of Earth Sciences, University of Lausanne, Lausanne, Switzerland
4. State Key Laboratory of Lithospheric and Environmental Coevolution, Institute of Geology and Geophysics, Chinese Academy of Sciences, Beijing, China

*Corresponding author (Fengchang Bu: fengchang.bu@unil.ch)

**Abstract**

The intact intervals interrupting joint and fault connections, known as rock bridges, have long been interpreted as crucial in resisting rock mass instabilities like faulting, landslides, and earthquakes. Despite extensive research on rock bridges across rock mechanics, tectonics, geohazards, seismology, and rock engineering, mainstream geoscientific thought treats rock bridges as conceptual entities rather than analysable objects. Following a manual review of each of the 2538 rock-bridge research spanning 1962–2025, we attribute this knowledge gap to the absence of a rock-bridge mechanics framework, which has resulted in domain impasses across rock-bridge definition, measurement, analytical method, resistance magnitude, failure mechanism, fracture morphology, stress-orientation dependence, determinant, and field incorporation. Recent advances in failure criteria for rock bridges promise to serve as the rock-bridge mechanics framework to address these impasses, as they clarify the relationship between rock-bridge configurations and mechanical behaviour, allowing us to analyse rock bridges in the associated geological phenomena and rock engineering, such as landslide triggering, fault linkage, and earthquake termination.

## 1. Introduction

Discontinuities are the most common result of brittle and semi-brittle fracture of crustal rocks (Müller, 1962; Price, 1966; Jaeger et al., 1969). The International Society for Rock Mechanics and Rock Engineering (ISRM) (1978) defined discontinuity as a general term for any mechanical discontinuity in a rock mass having zero or low tensile strength, as well as a collective term for most types of joints, weak bedding planes, weak schistosity planes, weakness zones and faults. Discontinuities profoundly affect the strength, deformation, permeability, and stability of rock masses, thereby playing a crucial role both in geological processes, including earthquakes (Brace and Bombolakis, 1963), tectonic deformation (Pollard and Fletcher, 2007), and mineralisation (Sibson, 1977), and in rock engineering applications, such as mining excavation (Brady and Brown, 1985), slope stability (Hoek, 2000), and dam foundation (Hudson and Harrison, 2000). Therefore, discontinuities have long been a central theme in geomechanics (Barton, 1976; Goodman, 1980; Bieniawski, 1989).

Given that natural discontinuities like joints and faults are often not fully continuous, Hudson and Priest (1983) classified discontinuities into persistent and impersistent types. ISRM (1978) quantified the characteristics of impersistent discontinuities as discontinuity persistence *K*, defined as the areal extent or size of a discontinuity within a plane (Fig. 1a):

$$K = \lim_{A_D \to \infty} \frac{\sum_i a_{D_i}}{A_D} \tag{1}$$

where $D$ is a region of the plane with area $A_D$, and $a_{Di}$ is the area of the $i$th discontinuity in $D$ (Einstein et al., 1983).

Field mapping of discontinuities relies primarily on outcrop surface traces; consequently, *K* is predominantly quantified by discontinuity trace lengths (Fig. 1b):

$$K = \lim_{L_S \to \infty} \frac{\sum_i l_{S_i}}{L_S} \tag{2}$$

where $L_S$ is the length of a straight line segment *S*, and $l_{S\,i}$ is the length of the $i$th joint segment in *S* (Einstein et al., 1983). Furthermore, Jennings (1970) divided $L_S$ into the discontinuity length $l_{Si}$ and rock-bridge length $R_i$, and *K* is expressed as (Fig. 1b):

$$K = \frac{\sum_i l_{S_i}}{\sum_i l_{S_i} + \sum_i R_i} \tag{3}$$

where a rock bridge is defined as a portion of intact rock separating discontinuities.

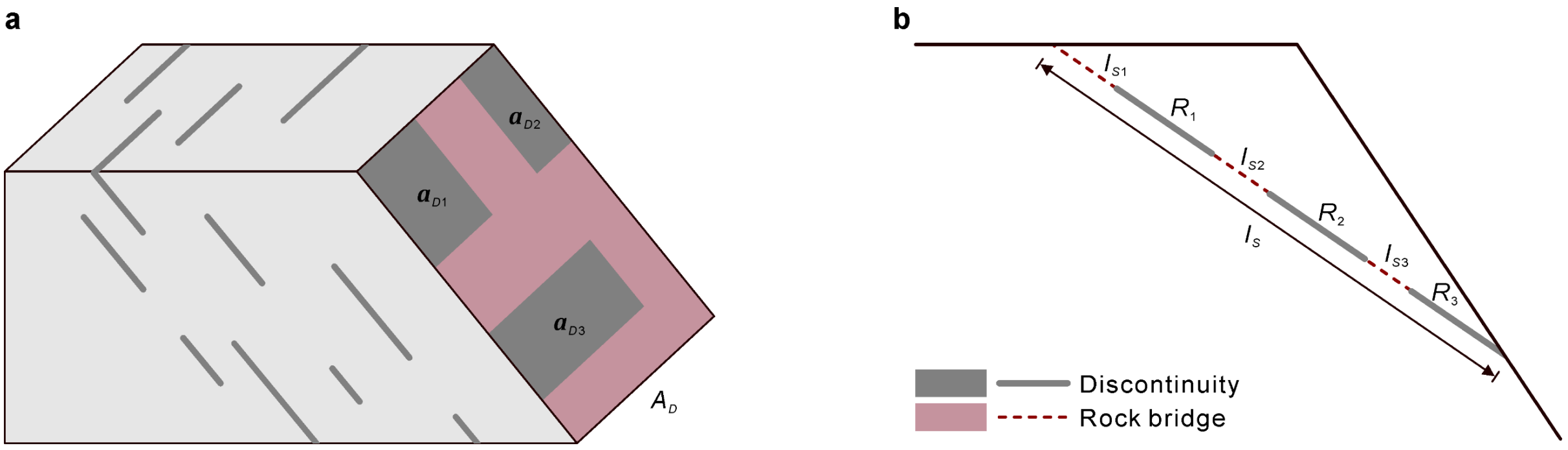

**Fig. 1. Discontinuity persistence. (a)** Persistence defined as area ratio, where $a_{D1}$, $a_{D2}$, and $a_{D3}$ denote the discontinuity areas, and $A_D$ denotes the total area. Modified from Müller (1962). **(b)** Persistence defined as length ratio area, where $l_{S1}$, $l_{S2}$, and $l_{S3}$ denote the discontinuity lengths, $R_1$, $R_2$, and $R_3$ denote the rock-bridge lengths, and $l_S$ denotes the total length. Modified from Einstein et al. (1983).

Discontinuities and rock bridges represent basic geometric features in geological mapping of rock masses. Any subsequent analysis of rock mass behaviour necessitates the integration of their mechanical parameters. For discontinuities, as they present zero or low tensile strength (ISRM, 1978), the integrity of the rock mass is significantly degraded, leading to a high susceptibility to instability (Hoek and Brown, 2019). The study of discontinuities boasts a long history and an extensive literature; although some challenges persist, the theoretical and technical frameworks have reached relative maturity, including their measurement (Hudson and Priest, 1983), statistics (Einstein and Baecher, 1983), strength (Barton and Choubey, 1977), deformation (Bandis et al., 1983), stability (Goodman and Shi, 1985), morphology (Xie, 1993), and permeability (Witherspoon et al., 1980).

For rock bridges, as the stiffness and strength of the rock matrix are significantly higher than those of discontinuities, rock bridges are always interpreted as zones of considerable stress concentrations and critical factors to resist rock mass failure (Segall and Pollard, 1980; Einstein et al., 1983; Wesnousky, 1988). Müller (1962) and Terzaghi (1962) pioneered the importance of rock bridges in slope stability. Exposed to natural events or human activities, intact rock bridges would accumulate damage to rupture, interconnecting intermittent or separate discontinuities and forming slip surfaces to trigger landslides (Fig. 2a,b), for example, the 1991 Randa rockslide (Stead and Eberhardt, 2013), 2007 Chamousset rockfall (Lévy et al., 2010), 2017 Pizzo Cengalo rock avalanche (Walter et al., 2020), and 2023 La Praz collapse (Bottelin and Baillet, 2024).

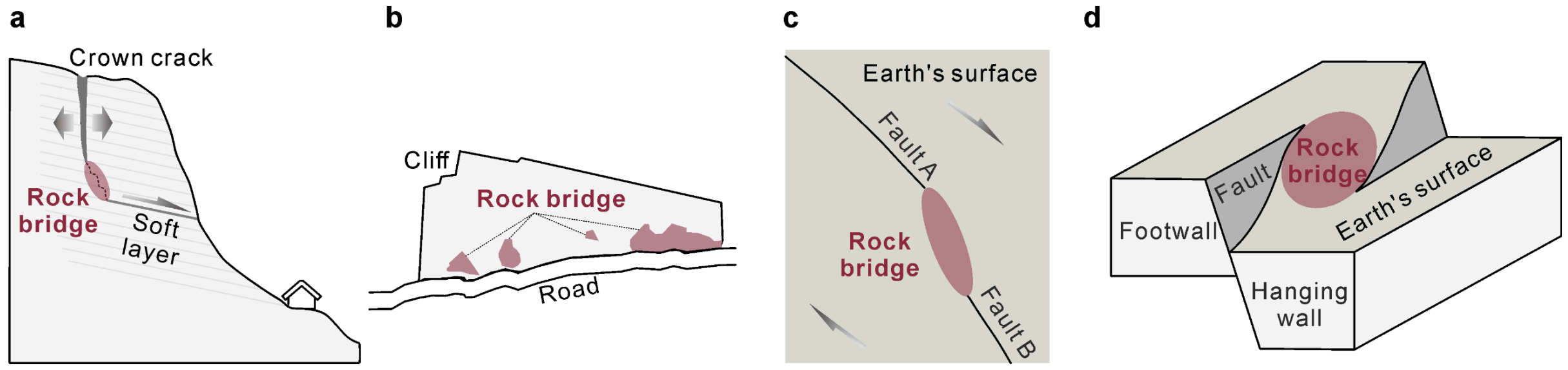

**Fig. 2. Schematic view of rock bridges across disciplines. (a)** A rock bridge resists a landslide. **(b)** Rock bridges exposed by cliff retreat. **(c)** Faults interact at a rock bridge. **(d)** A rock bridge transfers displacement in a normal fault.

In seismology, rock bridges are more broadly referred to as steps or stepovers (Fig. 2c), which may affect nucleation, propagation, and termination in earthquake ruptures (King and Nábělek, 1985; Sibson, 1985; Schultz, 1989; Harris et al., 1991; Wesnousky, 2006; Field et al., 2014; Scholz, 2019; Lee et al., 2024). In particular, stress concentrations induced by slip at active faults may stop ruptures at stepovers or penetrate these bridging structures to jump the ruptures to other faults (Harris and Day, 1999; Wesnousky, 2006; Oglesby, 2008), such as rupture jumps in the Kaikōura earthquake (Hamling et al., 2017), Ridgecrest earthquake (Ross et al., 2019), and El Mayor–Cucapah earthquake (Oskin et al., 2012).

In structural geology, rock bridges are usually termed relay and transfer structures (Fig. 2d), where faults interact, and strain or displacement is transferred or relayed from one structure to another (Fossen and Rotevatn, 2016). Their evolution governs fault nucleation, coalescence, and maturation, thereby affecting the associated geological phenomena (Peacock and Sanderson, 1991; Cartwright et al., 1995). For example, whether a rock bridge can resist fault hard-linkage determines the seismogenic potential, as the rupture length and slip magnitude differ markedly before and after physical connections (Walsh et al., 2003; Hodge et al., 2018; Lee et al., 2024). As rock bridges develop exceptional damage zones, they always facilitate cross-stratal fluid flow, controlling the distribution of crustal fluids, sequestered $CO_2$, and magmatic-hydrothermal mineralisation (Anders and Schlische, 1994; Bense and Van Balen, 2004; Fossen and Rotevatn, 2016).

Despite the abovementioned importance of rock bridges, research on rock bridges is much more limited than that on discontinuities. Especially, the absence of a systematic review on rock-bridge research has led to significant knowledge gaps concerning the definition, field incorporation, measurement, analytical method, and understanding of the mechanical behaviour of rock bridges like resistance magnitude, failure mechanism, fracture morphology, and stress-orientation dependence, as well as contemporary challenges. For instance, in rock engineering, rock bridges suffer from limited recognition, with few engineers incorporating them into engineering designs; in the laboratory, numerous redundant yet questionable experiments remain. In this context, through a comprehensive historical retrospection in the rock-bridge domain, this review interrogates the origin of these knowledge gaps, proposes strategies for transcendence, and explores the subsequent benefits.

## 2. Data collection and overview

A Google Scholar search for the term "rock bridge" spanning 1962–2025 yields 5510 publications, where years of record are determined by official publication instead of by preprint or online dates. Following manual screening to exclude irrelevant entries, duplicates, and instances where the term appears only in references, 2538 valid publications remain. As shown in Fig. 3a, the volume of research on rock bridges has increased annually, delineating four phases based on landmark publications:

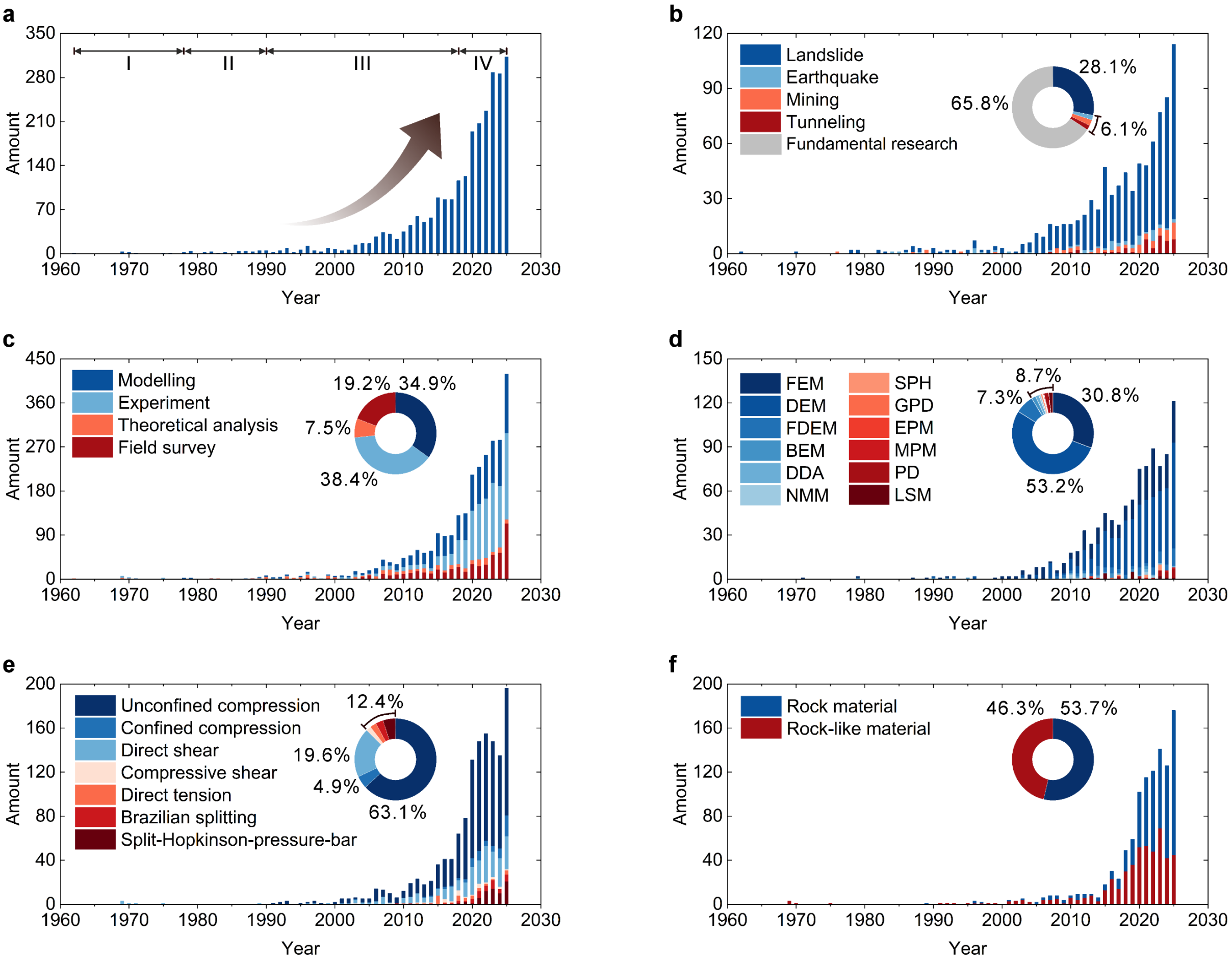

**Fig. 3. Rock-bridge research landscape. (a)** Volume. **(b)** Domain. **(c)** Methodology. **(d)** Modelling. Full names of abbreviations are provided in the main text. **(e)** Experiment. **(f)** Experimental material.

I Conceptual genesis and theoretical grounding (1962–1970). In the description of discontinuities, Müller (1962) conceptualised the interval between tips or edges of adjacent impersistent discontinuities as "material bridge" (termed "Materialbrücken" in German). Terzaghi (1962) conceptualised this entity as "gap", a possible resistance locus for rock slope instabilities. Both "material bridge" and "gap" are the conceptual embryonic forms of "rock bridge". Lajtai (1969a; 1969b; 1969c) pioneered the theories for deriving the strength of rock bridges in direct shear, with experimental validation by introducing direct shear tests on rock bridges. By refining the calculation of rock mass cohesion considering rock bridges from Terzaghi (1962), Jennings (1970) established a formal mathematical expression of rock mass strength, namely, Jennings's criterion, rendering the practical analysis of rock bridges feasible. Thus far, the paradigms established by Müller, Terzaghi, Lajtai, and Jennings remain widely utilised.

II Practical implementation (1970–1990). Following the abovementioned pioneering work, the importance of rock bridges has gained traction across disciplines, and the integration of rock bridges into engineering practice and geological research has emerged. For instance, during the construction of the Auburn Dam in California, USA, Von Thun and Tarbox (1971) incorporated 10% of the observed rock bridges into the finite-element analysis of the dam foundation, as well as the inaugural numerical simulation of rock bridges. In the design of the Marble Peak underground mine in Arizona, USA, Nicholas (1976) considered rock bridges in the step-path failure of mine pillars. The findings were later instrumentalised through a computer model designed to evaluate the step-path failure in open-pit slopes (Call and Nicholas, 1978), which have long been an important content in the stability analysis of rock slopes (Barton, 1971; Jaeger, 1971; Stimpson, 1978). In particular, Einstein and Baecher (1983) introduced probabilistic and statistical methods to investigate the effect of rock bridges on rock slope stability, constituting essential practical guidance for analysing rock bridges in rock slopes. Given the traction of rock bridges in dam foundation, mining excavation, and rock slope stability, researchers have sought to incorporate rock bridges into rock mass classification systems. When characterising the "Lucanian Dolomites" sandstone outcrop in Basilicata, Italy, Alexander and Rendell (1986) noted that "the neglected factor in most cases is the presence of rock bridges, or continuous rock that spans the basal joints". In the mining rock-mass rating (MRMR) classification system by Laubscher (1990), rock-bridge percent is determined by the fracture frequency rating (FFR) and an empirical constant, marking the relative maturation of the practical implementation of rock bridges. In addition to the engineering practice, a parallel surge in research on rock bridges has emerged within the geological domain. Earlier field investigations had demonstrated that faults are discontinuous geological features consisting of numerous discrete segments, always featuring en échelon arrays (Ramsay, 1968; Deng and Zhang, 1984). Nicholson and Pollard (1985) established a correlation between the linkage of en échelon faults and the rock-bridge failure, elevating rock bridges as an important topic in fault mechanics (Segall and Pollard, 1980). Furthermore, by establishing the relationship between rock-bridge width and earthquake rupture length of major strike-slip fault zones in California, USA, and Turkey, Wesnousky (1988) concluded that rock bridges can impede or arrest the propagation of earthquake ruptures. Since then, rock bridges have emerged as a pivotal subject within the field of seismology.

III Methodological maturation (1990–2018). Following the abovementioned foundational contributions, rock-bridge research entered a prolonged phase of incremental advancement. Driven by rapid technological progress, a diverse array of methodologies has since been integrated into the study of rock bridges. The scope of research encompasses landslides, earthquakes, mining, tunnelling, and fundamental studies, utilising methodologies such as modelling, experiment, theoretical analysis, and field survey. Investigations into rock bridges have transitioned towards diversification. In the landslide domain, hundreds of case studies had shown field evidence of rock bridges, covering the measurement, strength, failure mode, and cracking process. Extensive experimental and numerical studies had incorporated prescribed rock bridges to establish slope models to investigate their mechanical behaviours. In other rock engineering domains, the boundaries of rock-bridge research had also been expanded; for example, Diederichs and Kaiser (1999) established a model using crack and rock-bridge analogues to investigate the failure mechanisms in mining excavation. Compared with rock engineering, geological research on rock bridges relies more heavily on empirical observation and statistical analysis. In fault mechanics, geodetic and statistical analyses had revealed the scaling-law relationships between displacement and fault length, separation, and overlap to establish fault linkage criteria (Cartwright et al., 1995; Soliva and Benedicto, 2004). In the seismology domain, statistical and numerical analyses point to a threshold length of 4–5 km for rock bridges in arresting seismic rupture propagation (Harris and Day, 1999; Duan and Oglesby, 2006; Wesnousky, 2006). This phase also witnessed the emergence of a substantial body of fundamental research on rock bridges. As rock-bridge failure indicates the coalescence of pre-existing discontinuities, numerous theoretical, experimental, numerical, and field-based studies investigated rock-bridge failure by prescribing discontinuities with various arrangements within rock or rock-like materials in diverse stress regimes. Advancements in monitoring techniques, such as acoustic emission (AE), digital image correlation (DIC), X-ray computed tomography (CT), and thermal infrared imaging, had enabled a more granular investigation of the failure morphologies, failure mechanisms, and influencing factors of rock bridges. Distinct physical field responses of rock bridges, including those under dynamic loading, cyclic loading, freeze-thaw cycles, and wetting-drying cycles, had been captured across various conditions.

IV Expansion alongside debate (2018–2025). Benefitting from the abovementioned methodologies, research on rock bridges is experiencing explosive growth, and is becoming increasingly granular and multifaceted. This trend is attributable both to the heightened emphasis placed on research into rock bridges and, to some extent, the information explosion and the advent of artificial intelligence (AI). Taking landslide research as an example, the frequency of such disasters has intensified in recent years due to accelerating climate change. Numerous landslides have been confirmed to result from rock-bridge failure, leading to a heightened focus on this specific type. Concurrently, new technological tools have identified rock-bridge failure as the primary driver in several significant landslides, further bolstering interest in these features. Behind this rapid expansion, underlying challenges are emerging. Elmo et al. (2018) proposed six challenges in the characterisation of rock bridges in rock slopes, among which the first is “what is a rock bridge and what parameters govern whether a given intact rock portion of the rock mass can or should be defined as a rock bridge”, that is, the conceptual validity of rock bridges. Within the seismic domain, scholars have recognised that statistical correlations between rock-bridge length and earthquake termination are insufficient; facilitated by advancements in numerical simulation, recent research has notified multiple influencing factors, such as fault geometry (Mia et al., 2024), friction (Ryan and Oglesby, 2014), and interaction (Yin et al., 2023), initial stress (Lozos and Harris, 2020; Wang et al., 2020a), topography (Douilly, 2023), and poroelasticity (Huang et al., 2025). Additionally, despite the vast accumulation of experimental and simulation data on the physical responses of rock bridges at a laboratory scale, the strength of rock bridges is highly puzzling, the cracking process is highly complicated, and the failure morphology is highly intricate. Can we break through the boundaries of phenomenon-based observation to advance our fundamental understanding of rock-bridge mechanics?

The preceding overview of the four phases underscores that rock-bridge research is interdisciplinary and multi-methodological. To provide an overview of these diverse fields and approaches, the 2538 valid publications are categorised into several groups, as shown in Fig. 3b–f.

As illustrated in Fig. 3b, fundamental research into rock bridges predominates at 65.8%, encompassing studies devoid of a particular engineering or geological context despite possible contextual mentions in the introduction. Landslides represent the second-largest share and the leading application domain at 28.1%, suggesting that rock bridges remain primarily studied within this context. The remaining sectors comprising earthquake, mining, and tunnelling account for a combined 6.1%, with individual shares of 1.9%, 2.3%, and 1.9%, respectively. Notably, as Fig. 3b is based solely on publication volume, it may not fully reflect the relative popularity of rock bridges across fields, given the varying sizes of their respective research communities.

Figure 3c presents the classification of research methodologies, illustrating that modelling and experiment are the primary approaches in rock-bridge research, accounting for 34.9% and 38.4%, respectively. Field investigations also show a significant presence at 19.2%, while theoretical analysis represents the smallest share at only 7.5%. The annual distribution reveals that most methodologies have experienced general growth, with the exception of theoretical analysis, which has remained stagnant or even declined slightly.

Figure 3d further categorises the modelling methods, illustrating a scope that encompasses finite element method (FEM), discrete element method (DEM), finite-discrete element method (FDEM), boundary element method (BEM), discontinuous deformation analysis (DDA), numerical manifold method (NMM), smoothed particle hydrodynamics (SPH), general particle dynamics (GPD), elasto-plastic model (EPM), material point method (MPM), peridynamics (PD), and lattice spring model (LSM). Among these, DEM and FEM emerge as the dominant approaches; DEM, in particular, accounts for a substantial 53.2%, representing the primary tool for rock-bridge simulation due to its superior capability in modelling fracturing processes. FEM, a long-established methodology, also maintains a significant position, contributing 30.8% of the total. As an emerging modelling technique, FDEM has seen a steady rise in publications, reaching 7.3%. The remaining methods account for 8.7%, many of which utilise rock-bridge simulations to validate algorithmic efficacy, whilst others explore novel aspects of the mechanical behaviour of rock bridges.

Figure 3e further categorises the experimental methodologies, which encompass the core suite of rock mechanics testing: unconfined compression, confined compression (including true triaxial compression), direct shear, compressive shear, direct tension, Brazilian splitting, and split-Hopkinson-pressure-bar test. Unconfined compression accounts for the largest share at 63.1%, owing to its simplicity. Given that rock bridges are frequently subjected to shear stress in landslide and fault contexts, direct shear testing is also extensively employed, representing the second-largest proportion at 19.6%. Recent technological advancements have facilitated the development of true triaxial testing apparatus, enabling confined compression to secure a 4.9% share. Other experimental methods collectively constitute the remaining 12.4%.

Figure 3f categorises the experimental materials. Prior to 2018, rock-like materials predominated in experimental studies, largely due to the technical difficulties associated with pre-fabricating rock bridges within natural rock samples. These analogues included cement, gypsum, polymethyl methacrylate (PMMA), and various composite mixtures. Since 2018, however, advancements in processing technologies have driven a steady increase in the use of natural rock, which now accounts for a cumulative 53.7% of the literature.

## 3. Definition

### 3.1. Conceptual genesis

The concept of a rock bridge originated from the description of discontinuities by Müller (1962):

*‘Ausgedehnte Klüfte begünstigen Gleitungen und Ausweichbewegungen, aber auch die Bildung von Quellröhren und Sickerwegen des Wassers usw. weit mehr, als viele Kluftindividuen der gleichen Schar von gleicher Gesamtfläche, zwischen denen „Materialbrücken“ bestehenblieben’.*

with the English version translated by Google Translate:

*‘Extensive joints facilitate sliding and evasive movements, as well as the formation of spring conduits and seepage paths for water, etc., far more than many individual joints of the same set with the same total surface area between which “material bridges” remain’.*

In the same year, a similar concept was introduced in a landslide study conducted by Terzaghi (1962):

*“Every body of rock contains more or less discontinuous joints with very different degrees of discontinuity. A section following a discontinuous joint cannot enter an adjacent joint without cutting across intact rock. Those portions of a section which are located in intact rock will be referred to as gaps”.*

In structural geology, a similar concept was introduced by Ramsay (1968):

*“The sectors of unfaulted rock lying between the en-echelon fault relays are termed fault bridges”.*

The “material bridge”, “gap”, and “fault bridge” are what we have subsequently come to know as the “rock bridge”. The documentation of the term “rock bridge” occurred in a fundamental study conducted by Lajtai (1969c):

*“In complex rock structures with unfavourably orientated planes of weakness directions, a developing failure surface will have to shift from one set of discontinuities to another, cutting across solid rock (rock bridges) in the process”.*

Thereafter, “rock bridge” increasingly became the predominant terminology. Various definitions have been proposed by several researchers. For example, Jennings (1970) defined it as the intact material between joints; Stimpson (1978) defined it as the intact rock in a discontinuously jointed rock mass; Ladanyi and Archambault (1980) defined it as the blocky rock masses without through-going discontinuities; Barton (1990) defined it as the intact portion between joints; Gehle and Kutter (2003) defined it as the region between adjacent discontinuities.

Kim et al. (2007a; 2007b) defined it as a small bridge of intact rock separating coplanar or non-coplanar discontinuities. Hencher and Richards (2015) defined it as the discrete section of intact rock or incipient discontinuities that have not yet fully developed as open fractures. Tuckey and Stead (2016) defined it as the interval of intact rock between the tips or edges of adjacent discontinuities. Elmo et al. (2018) defined it as a portion of intact rock separating joint surfaces. These findings allow for a concise definition of a rock bridge: A portion of intact rock between tips or edges of adjacent impersistent discontinuities. The conceptual genesis indicates that a rock bridge is a purely geometric concept derived from observation and devoid of any inherent mechanical connotations.

### 3.2. Conceptual development

The subsequent conceptual development has branched into two primary streams: disciplinary specialisation and the extension to mechanical connotations.

As the definition of rock bridges is associated with discontinuities, different types of discontinuities can be mapped to distinct terminologies of rock bridges. In rock engineering, “rock bridge” is more prevalent, likely because most rock engineering primarily deals with joints. In fault systems, compared to the “fault bridge” defined by Ramsay (1968), the term “step” or “gap” may be more frequently employed (Sibson, 1985). Biasi and Wesnousky (2016) defined steps as those sites where a rupture or fault trace is interrupted by a discontinuity that may be described as requiring an observer to step approximately orthogonally from the end of a rupture segment to find the continuation of the fault or rupture or their projections along strike, and defined gaps as those sites where mapping is considered adequate to recognise that there is an absence of surface rupture along a continuous fault trace that then resumes somewhere farther along strike. The more specific term for “step” is “stepover” (Aydin and Nur, 1985). Barka and Kadinsky-Cade (1988) defined stepovers as offsets in the fault trace. Regardless of the terminology employed across disciplines, the underlying essence remains consistent with the concept of rock bridges.

As rock bridges are always interpreted as zones of considerable stress concentrations and critical factors to resist instability, researchers have extended the geometric concept “rock bridge” to mechanical connotations. In the landslide domain, Zhang et al. (1981) conceptualised the structure along a slip surface with a considerable stress concentration and a large bearing capacity to resist instability as “locked segment” (similar terms include locked patch, locked section, locking section, etc.). Qin et al. (2010) and their colleagues (Xue et al., 2014; Chen et al., 2018) proposed a criterion to predict locked segment failure according to the weakest-link and renormalisation group theories. Geological investigations of numerous large-scale landslides have provided extensive empirical support for the concept of locked segments, such as the 1933 Diexi (Huang, 2012), 1943 Zana (Huang, 2012), 1971 Libby dam (Chen et al., 2018), 1980 Yanchihe (Chen et al., 2018), 1983 Saleshan (Huang et al., 2016), 1985 Xintan (Chen et al., 2021a), 1989 Xikou (Huang, 2012), 1991 Touzhai (Huang, 2012), 1996 Yinjiangyankou (Pan et al., 2017), 2000 Yigong (Huang, 2012), 2003 Qianjiangping (Zhang et al., 2018), 2008 Daguangbao (Cui et al., 2018), 2008 Wangjiayan (Hu et al., 2023), 2008 Zhengjiashan (Xu et al., 2015), 2008 Tangjiashan (Xu et al., 2013), 2009 Jiweishan (Yin et al., 2011), 2014 Xiaoba (Lin et al., 2018), 2015 Shanyang (Zhan et al., 2024), 2017 Xinmo (Xi et al., 2025), 2018 Baige (Cao et al., 2025), and Tizicao potential landslide (Zhou et al., 2022b), leading to its broader adoption. For the locked-segment-controlled landslides, there are two primary classification systems: Qin and his colleagues (Pan et al., 2014; Chen et al., 2018) classified them into inward-dipping cross-layer shear, outward-dipping cross-layer shear, bedding direct shear, homogeneous rock bridge, retaining wall, and supporting arch, determined by the spatial intersection of potential sliding surfaces with stratigraphic bedding, the slope-to-strata orientation, and the morphological characteristics of locked segments; while Huang et al. (2017) classified them into three sections type, retaining wall type, anti-slide block type, linear multilevel type, and stepped multistage type based on the existing form and the position of locked segments. Regardless of the classification system used, a rock bridge is the predominant type of a locked segment. Therefore, the concepts of a locked segment and a rock bridge are always used interchangeably without distinction (Al-E’Bayat et al., 2024; Diao et al., 2025). However, a rock bridge may be a locked segment, but a locked segment may not be a rock bridge. A rock bridge can only be designated as a locked segment if it strictly satisfies the definition of a locked segment. This necessitates an assessment of the resistance provided by a rock bridge.

Similar mechanical extension has even challenged the validity of how a rock bridge is defined. Elmo et al. (2007; 2013) defined rock bridges as complex zones of influence where failure is governed by fracture propagation and coalescence. The resulting critical path follows the orientation of minimum shear resistance, highlighting that the effective rock bridge is a combined product of geometry, geomechanical parameters, and applied field stresses rather than a simple geometric distance. Indeed, while our primary concern is the mechanical role of a rock bridge, such a definition diverges from the core nature of a rock bridge as a purely geometric construct, which is inherently devoid of mechanical connotations.

In fault systems, stepovers (rock bridges) can be further classified by stress states and by geometry. In left-lateral vertical strike-slip systems, concurrent slip results in restraining (compression) at right steps and releasing (dilation) at left steps, namely, restraining and releasing stepovers. This relationship is reversed for right-lateral faults. Geometrically, the stepover width is defined as the perpendicular distance between fault segments, while the overlap represents their along-strike crossover distance. Accordingly, Barka and Kadinsky-Cade (1988) classified stepovers into offset, separation, and overlap. As stepovers are the discontinuous parts in a fault system, they may transfer strain or displacement among faults. In normal faults, Fossen and Rotevatn (2016) defined the locations of fault interaction where strain or displacement is transferred or relayed from one structure to another as relay and transfer structures. Therefore, in some studies, relay/transfer structures are interchangeable with stepovers. Nevertheless, a relay/transfer structure can be a stepover, but a stepover may not be a relay/transfer structure. Whether a stepover can be referred to as a relay/transfer structure should consider the mechanical behaviour of the stepover.

### 3.3. Conceptual challenge

The above review of rock-bridge concept reveals two primary challenges: disciplinary specialisation and conceptual extension. For the disciplinary specialisation, terminological inconsistency across fields, such as "rock bridge" in rock engineering and "stepover" in seismology. Even in the same field, terminological inconsistency remains; for instance, "fault bridge", "stepover", "step", "gap", "rock bridge", and "relay structures" are always used interchangeably in fault mechanics (Ramsay, 1968; Aydin and Nur, 1985; Biasi and Wesnousky, 2016; Prastyani et al., 2025). This terminological inconsistency may create disciplinary silos that hinder knowledge exchange.

The other challenge is that the mechanical extension of a rock bridge remains poorly defined. While often intuitively equated with "locked segment" or "relay structure", "rock bridge" is a geometric construct whereas "locked segment" or "relay structure" is mechanical. Determining whether "rock bridge" functions as a mechanical construct requires clarifying its mechanical properties, particularly its resistance against instability; failing to do so risks conceptual misunderstandings and undermines the perceived significance of rock bridges, as not all instances of rock bridges are functional.

Additional challenges further complicate the concept of rock bridges, as detailed in subsequent sections.

## 4. Field incorporation

The importance of rock bridges also originates from the description of discontinuities by Müller (1962):

*"Entstehung welliger Scherklüfte bei großem Deformationsausmaß infolge Durchreißens der*

*Materialbrücken".*

with the English version translated by Google Translate:

*"Formation of wavy shear joints at high degrees of deformation due to the tearing of material bridges".*

At the first ISRM Congress in 1966, Müller (1966) more explicitly documented the importance of rock bridges in his presentation entitled "The progressive failure in jointed media":

*"In discontinues the mass strength is composed of genuine shearing strength of the material bridges and of friction in the joints. the deformation resistance of the material bridges takes effect at much smaller deformations than the joint friction does".*

The significance of rock bridges was subsequently extended across various domains of engineering practice and geological investigation.

### 4.1. Rock bridge in landslides

In a landslide study conducted by Terzaghi (1962), the importance of rock bridges was noted as:

*"A shear failure along the section is resisted by both pressure-conditioned shearing resistance and the cohesion of the rock located in the gaps between joints".*

From the observation of landslides, Jennings (1970) concluded that:

*"Observation shows that failure of a slope in unweathered rock tends to take place on planes which pass partly along pre-existing joint surfaces and partly through intact rock material. For this reason the stability of the slope depends to a large extent on the degree of mobilization of the intact rock strength i.e. on the measure of interlocking of the jointed material of the slope."*

From that point onwards, the significance of rock bridges in landslides became increasingly recognised (Robertson, 1970; Einstein et al., 1983; Sjöberg, 1999; Stead and Eberhardt, 2013). Stimpson (1978) demonstrated that only a minor proportion of the surface area needs to consist of a rock bridge to ensure stability against joint-rock-bridge planar sliding. In a stability analysis of rock slopes, Martin (1978) concluded that the occurrence of less than 10% rock bridges could provide sufficient resistance along a potential failure surface for slopes more than 300 m high in moderately hard rock. As the understanding of rock bridges has deepened, this critical threshold has undergone continuous refinement. For example, Havaej et al. (2012) demonstrated that 2% rock-bridge content was sufficient to stabilise an unstable pentahedral rock slope wedge. Tuckey and Stead (2016) noted that a small content of rock bridges ranging from 1% to 3% can significantly increase the slope stability. Elmo et al.

(2018) recognised that even a 5% inclusion of rock bridges stabilises carbonate cliffs. See Section 5.2 for more rock-bridge percent estimation. On the other hand, some researchers have highlighted the significance of rock bridges by quantifying the specific increments in stability they provide. For instance, Nilsen (1979) demonstrated that an increase of rock-bridge percent from 0 to 10% could yield a five-fold increase in the factor of safety from 0.3 to 1.7. Paronuzzi et al. (2016) demonstrated that even a minimal presence of rock bridges can significantly enhance the shear strength of a weak surface by 25–40%. Jbara and Tsesarsky (2025) concluded that neglecting rock bridges in hazard assessments would result in safety factor analyses that underestimate the long-term stability of centennial- to millennial-scale limestone cliffs.

Extensive field investigations have also demonstrated the importance of rock bridges in landslides. Huang (2012; 2015) analysed the mechanisms of large-scale landslides in western China in the 20th century, such as 1933 Diexi, 1943 Zana, 1980 Yanchihe, 1989 Xikou, 1991 Touzhai, and 2000 Yigong landslide. He concluded that rock-bridge failure is one of the main mechanisms of these landslides. The brittle failure of rock bridges under shear stress typically triggers high-speed and long-runout rockslides, driven by the significant disparity between peak and residual strengths of rock bridges. At the regional scale, following a detailed field investigation of twenty-five rockfalls in the French Subalpine Ranges, Frayssines and Hantz (2006) analysed the structural conditions of the rock masses, morphology of the initial cliff surface and the scar, and possible failure mechanisms and processes. They suggested that the dominant consideration for rockfall can be the proportion of rock bridges in the potential failure surface. Their further study conducted a limit equilibrium analysis to understand the roles of rock bridges in sliding and toppling (Frayssines and Hantz, 2009). They found that a very small rock bridge can provide significant resistance to rockfalls. Another regional field investigation of rock bridges is from de Vilder et al. (2017), who conducted a detailed field investigation of multiple rockfall scars on the North Yorkshire coast in the UK and measured the rock-bridge percent, shape, and locations by characterising failure scars. They found that larger instabilities are characterised by fracture propagation through multiple rock-bridge networks, whereas smaller failures typically involve the rupture of a single, critical bridge. Furthermore, the spatial distribution of rock bridges relative to the incipient rockfall geometry dictates the eventual failure mode. The observation of weathering along both discontinuity surfaces and previously ruptured rock bridges highlights the sequential nature of progressive rockfall detachment.

Here we select several representative landslide cases analysing rock bridges. For the 1991 Randa rockslide in Switzerland, Eberhardt et al. (2004a) conducted FDEM simulations and attributed the evolution of the rockslide to the successive rupture of rock bridges. They concluded that massive rock slope instability inherently requires the progressive degradation and destruction of the cohesion of rock bridges to trigger the catastrophic failure. Accordingly, they conceptualised massive rock slope sliding mechanisms as controlled by progressive shear plane development and internal rock mass deformation/damage (Fig. 4) and highlighted the importance of rock bridges in these processes. Furthermore, Gischig et al. (2011) characterised the rock mass structure and displacement patterns using photogrammetry, terrestrial laser scanning (TLS), ground-based differential interferometric synthetic aperture radar (GB-DInSAR), and 3D geodetic measurements. Above the basal sliding surface of May 1991, a distinct kinematic transition is evident: progressing from basal sliding to stepped planar failure, through a zone characterised by rock bridges, and ultimately culminating in tensile rupture adjacent to the vertical head scarp. Their further DEM simulations indicate that certain rock bridges must break prior to catastrophic failure.

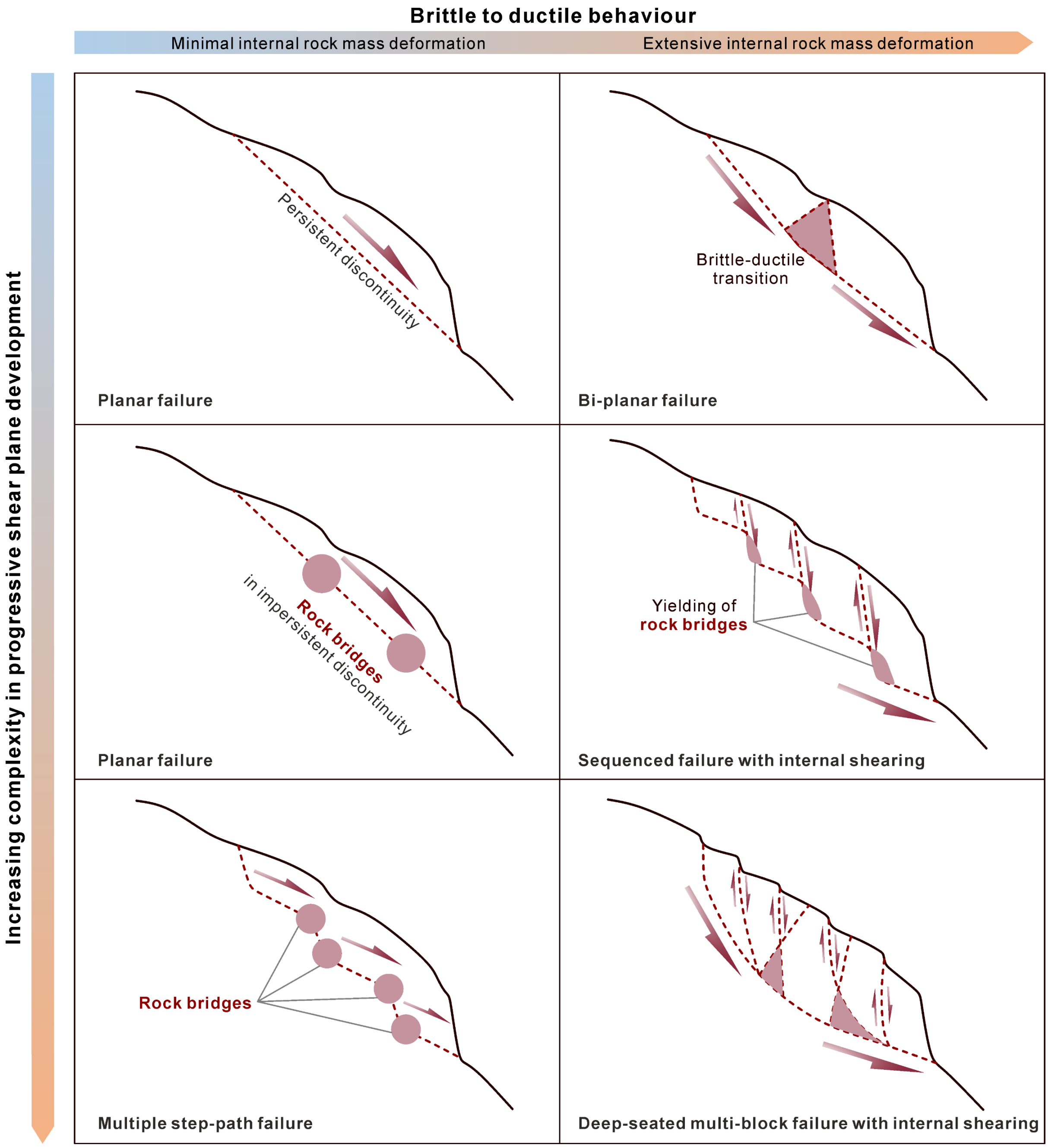

**Fig. 4. Mechanisms of massive rock slope instability**, modified from Eberhardt et al. (2004a).

Another representative case is the Åknes potential rockslide in Norway. Grøneng et al. (2009) interpreted a rock bridge with a percent of 1–3% based on geological mapping, geophysical surveys, and drilling of boreholes. They estimated the shear strength of the potential sliding surface as 0.5–0.75 MPa for a depth of 25 m and 1.05–1.3 MPa for a depth of 60 m. They also concluded that a rock bridge significantly contributes to the slope stability. Their further study highlighted the need for monitoring the disintegration of the rock bridge (Grøneng et al., 2011). Jaboyedoff et al. (2011) and Oppikofer et al. (2009; 2011) further conducted airborne laser scanning (ALS), TLS, and structural analysis to interpret the distinct displacement patterns of different rockslide subdomains. They concluded that the basal sliding surface was initially interrupted by rock bridges in the fold hinge zones, which had to yield during the progressive failure stage to create a continuous and stepped basal sliding surface.

Both the 1991 Randa rockslide and the Åknes potential rockslide are “stepped failure”, which is an important failure mechanism in rock slopes. The interaction between pre-existing discontinuities and the brittle failure of rock bridges can facilitate step-path mechanisms, or lead to complex sliding and toppling failures where the rupture of rock bridges provides the necessary sliding or release surfaces (Fig. 5). Jennings (1970) incorporated rock bridges into the limit equilibrium methods to calculate the shear strength of the stepped sliding surface and the factor of safety. Furthermore, Call and Nicholas (1978) developed a computer program to analyse the step-path failure, and their modelling gave the distributions of both step path angles and rock-bridge percent, which were used in the slope stability analysis at the Cyprus Bagdad mine, the Bamangwato Concessions’ Pikwe mine, and three cases in Idaho (Ristau, 1994). Moreover, McMahon (1979) introduced Monte Carlo simulations to simulate step-path failure, which was further integrated into the STEPSIM4 code by Baczynski (2000). Karami et al. (2007) compared DEM-simulated and limit equilibrium-derived results, and they stated that rock-bridge analysis is of critical significance in deep open-pit mines, where elevated *in situ* stresses at the slope toe can trigger the progressive rupture of rock bridges. Such stress-driven degradation facilitates the coalescence of discontinuities, ultimately leading to the development of a continuous step-path failure surface. Yan (2008) elaborated on traditional scanline mapping, digital photogrammetry, and 3D laser scanning techniques, along with numerical simulations, to investigate the resistance role of rock bridges in step-path failure. See Scavia (1995); Huang et al. (2015); and Gu et al. (2025) for more fundamental research on rock bridges in stepped failure in rock slopes.

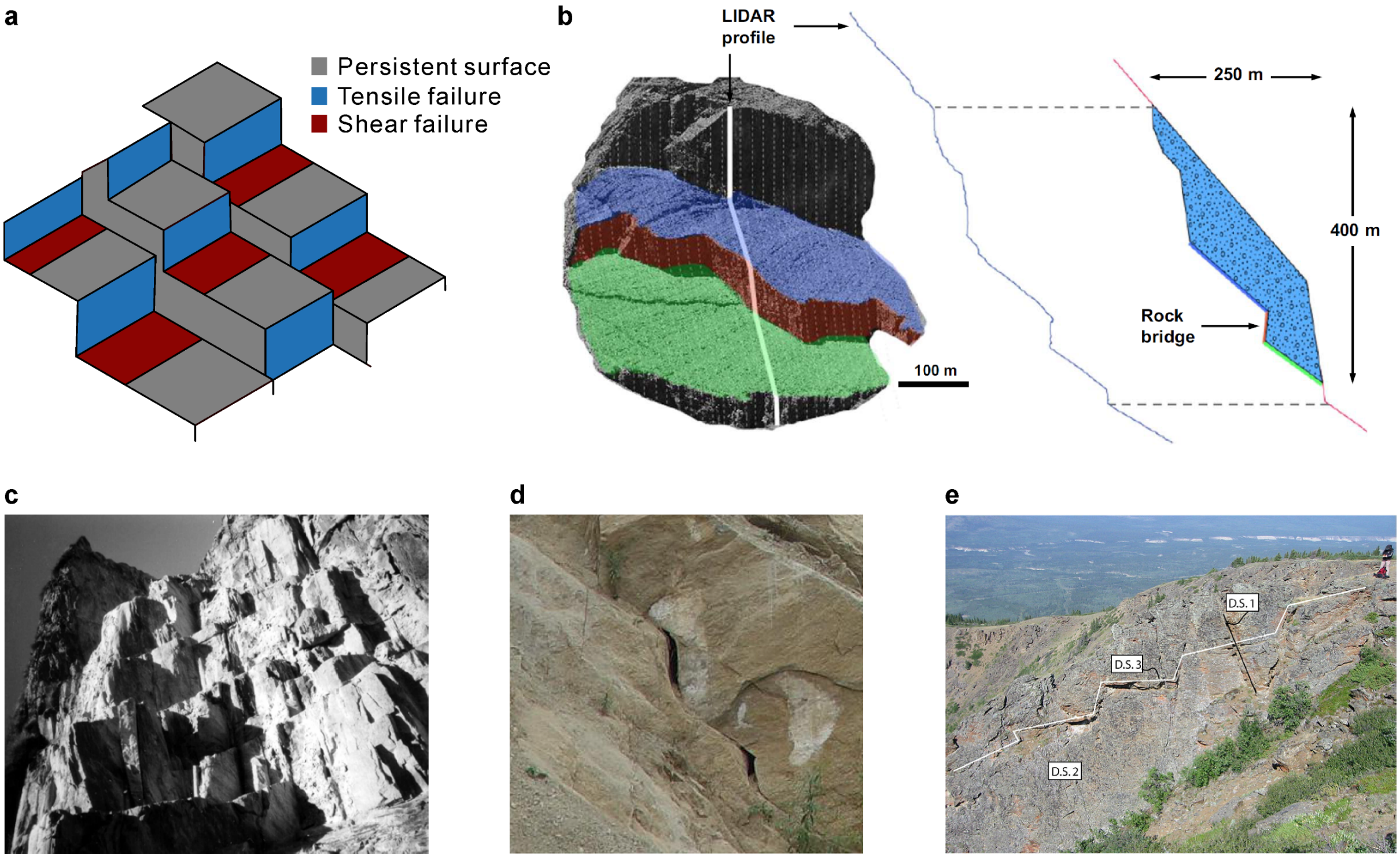

**Fig. 5. Stepped failure in landslides. (a)** Conceptual model of stepped failure, modified from Jennings (1970). **(b)** Stepped failure in the Cerro Catedral rockslide (García et al., 2018). **(c)** Stepped failure in the 1991 Randa rockslide (Eberhardt et al., 2004b). **(d)** Stepped failure in the reservoir slopes of the Xiaowan hydroelectric station (Huang et al., 2015). **(e)** Stepped failure in the Aishihik River landslide (Brideau et al., 2009).

Further documented rock-bridge controlled landslide cases include the 1983 Rosandra wedge collapse (Paronuzzi et al., 2015; Paronuzzi et al., 2016), 1991 Cellina rockfall (Paronuzzi and Serafini, 2009), 2004 Rosandra rockslide (Paronuzzi and Bolla, 2022), 2004 block caving-induced failure of the North wall at the Palabora Mine (Vyazmensky et al., 2010), 2007 Chamousset limestone column collapse (Lévy et al., 2010), 2008 Half Dome rockfall (Stock et al., 2011), 2009 Rhombus rockfall (Stock et al., 2012), 2011 Néron rockfall (Bonilla-Sierra et al., 2015), 2014 Xiaoba rockslide (Lin et al., 2018), 2019 Hüttschlag rockfall (Gerstner et al., 2023), 2017 Pusa collapse (Liu et al., 2025), 2023 Matterhorn rock pillar failure (Weber et al., 2025), Frank rockslide (Yan, 2008), Palliser rockslide (Sturzenegger and Stead, 2012), Drus potential collapse (Matasci et al., 2015), Diavik A154S rockfall (Sampaleanu, 2017), Cerro Catedral rockslide (García et al., 2018), Tizicao landslide (Zhou and Zhao, 2024), Hochvogel potential rock slope failure (Leinauer et al., 2024), KV42 Kings rock column potential failure (Alcaino-Olivares, 2024), Quincinetto landslide (Carlà et al., 2024), Ramon pillar potential failure (Jbara and Tsesarsky, 2025), etc.

In addition to the above field-based research, extensive efforts have also been made through numerical simulations to investigate the mechanical behaviour of rock bridges in landslides, and results reveal that the mechanical behaviour of rock bridges is controlled by multiple factors and is thus highly complex. Through 3D DEM simulation, Bonilla–Sierra et al. (2015) found that, as the sliding surface inclination increases, the failure mechanism of rock bridges transits from shear to tension; however, using LSM simulation, Al-E'Bayat et al. (2024) revealed that the failure mode is highly sensitive to discontinuity dip angle, transitioning from tension to shear–tension as the inclination increases. Further analysis of parameters including dip angle, spacing, and rock-bridge length indicates that impersistent joints constrain the extent of mobilisation zones compared to fully persistent systems. While wider spacing and longer rock bridges both improve the factor of safety and modify the depth of the failure zone, the degree of joint trace overlap is found to be statistically insignificant regarding slope behaviour.

The abovementioned field and numerical studies demonstrate the importance of rock bridges in the evolution of rock slopes. The mechanical behaviour of rock bridges is highly complex, modulated by topography, lithology, hydrogeology, tectonics, and slope structure, yet a unified mechanical paradigm remains elusive.

### 4.2. Rock bridge in other rock engineering

In dam foundations, rock bridges are always considered important factors for stability. During the construction of the Auburn Dam in California, USA, Von Thun and Tarbox (1971) incorporated 10% of the observed rock bridges into the finite-element analysis of the dam foundation. They concluded that a very few rock bridges would cause a substantial increase in the overall shear modulus of rock masses. For the Three-Gorges Dam foundation, Liu et al. (2003) conducted field investigations and geomechanical tests to investigate rock-bridge behaviour. Despite the density of gently dipping fractures, the rock mass remains structurally constrained by rock bridges that prevent the formation of a unified sliding surface. Dam instability is therefore contingent upon the rupture of rock bridges, which is necessary to create a continuous basal shear plane.

In underground mining, rock bridges also attract significant attention. Rock bridges are ubiquitous in mining pillars and surrounding rocks. Diederichs (1999) demonstrated the critical role of residual tensile strength, provided by rock bridges, in maintaining the stability of both gravity roof wedges and roof beams. His analysis further addressed the impacts of excavation-induced confinement loss, free-boundary interactions, and localised residual stresses stemming from material and geometric heterogeneities. Diederichs and Kaiser (1999) proved that the capacity of 1% rock bridges in hard rocks (unconfined compressive strength > 200 MPa) matched the capacity of a single cable bolt at least.

In open-pit mining, rock bridges also play an important role in blasting operations. LeBaron (2011)

conducted three blasts through an experimental approach to assess the effect of rock bridges on fragmentation. He found that the discontinuity persistence $K$ is directly related to the fragmentation degree. Aliabadian and Sharafisafa (2014) conducted DEM simulations on blast-wave propagation through rock bridges. Results indicate that stress concentration at rock bridges drives excessive fracturing, particularly during free-face reflection. Ultimately, the orientation and configuration of these impersistent joint systems significantly dictate the resulting rock fragmentation patterns.

### 4.3. Rock bridge in fault systems

The importance of fault bridges (rock bridges) is clearly established in the textbook *Folding and Fracturing of Rocks* by Ramsay (1968):

*"Because of the development of a strong differential bending of the fault bridges stresses may build up within the bridge and lead to a secondary failure across the bridge or to a sideways propagation of one or both tops of an en-echelon fault pair into the adjacent fault."*

Field descriptions by Burchfiel and Stewart (1966) constitute an early investigation into stepovers (rock bridges). They observed a pull-apart basin between right-lateral faults which step to the right in Death Valley, California, USA. Since then, the identification of fault stepovers has gained widespread attention. Aydin and Nur (1985) conducted a comprehensive review on stepovers in strike-slip tectonics. They classified stepovers as those along the strike and along the dip of faults. As exemplified by the along-strike stepover, if its sense is consistent with the fault slip, a pull-apart basin forms (Fig. 6a); if opposed, a push-up range forms (Fig. 6c). The pull-apart basin and push-up range correspond to the releasing and restraining stepovers, respectively. They also discussed the origins of stepovers, including the bending of initially straight faults, weak zones oriented slightly off the local plane, and curvature on fault traces. The importance of stepovers was explicitly listed as: (1) producing pull-apart basin and push-up ranges, which are potential oil, gas, and geothermal exploration targets; (2) enhancing permeability of fault zones by pull-apart openings or tail cracks along the faults, thus facilitating the formation of hydrothermal ore deposits; (3) generating heterogeneities along faults and thus controlling seismicity in active fault environments; (4) localising secondary structures, such as, joints, dikes, volcanoes, folds, cleavages, and dip-slip faults associated with strike-slip faulting; (5) enhancing rock rotations within the stepover area; and (6) controlling the location of back-arc extension and subduction-related volcanism.

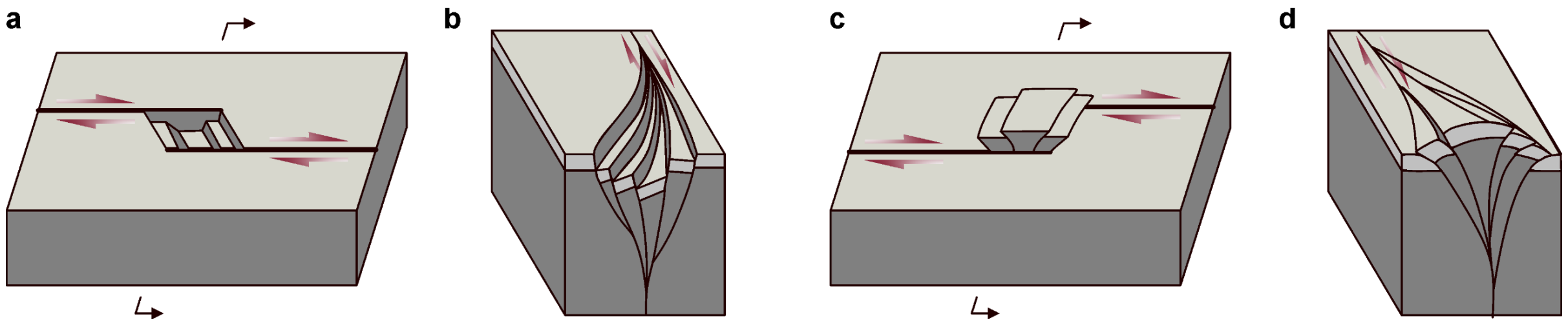

**Fig. 6. Stepover (rock bridge) in strike-slip faults. (a)** Releasing stepover (rock bridge) failure forms a pull-apart basin. **(b)** Cross-section view of the pull-apart basin is the negative flower structure. **(c)** Retaining stepover (rock bridge) forms a push-up range. **(d)** Cross-section view of the push-up range is the positive flower structure.

A compelling importance of stepovers within faults is their capacity to arrest earthquake rupture propagation. Wesnousky (2006) and Biasi and Wesnousky (2016) conducted a statistical analysis of moderate-to-strong earthquakes, and found that ~60–70% of them are terminated at stepovers. Early research in this field dates back to the 1980s. Segall and Pollard (1980) considered fault interaction in their 2D quasi-static analysis of strike-slip faults. They posited that restraining stepovers induce a significant rise in both mean and normal stress. Consequently, the restraining stepovers function as mechanical barriers to rupture propagation, effectively arresting earthquakes. In contrast, Sibson (1985) considered pore-fluid pressure, and concluded that the releasing stepovers would delay or terminate ruptures, as the decrease in fluid pressure in the tensile stress regime would increase fault strength. To clarify this debate, Harris et al. (1991) conducted 2D FEM simulations, and they concluded that releasing stepovers cause a time delay, leading to a lower apparent rupture velocity than restraining stepovers. The rupture is capable of jumping wider releasing stepovers than restraining stepovers. Their further simulation validated that the effect of the rupture on the undrained pore fluids is to inhibit the rupture from jumping releasing stepovers (Harris and Day, 1993). Oglesby (2005) provided valuable geomechanical insights into this topic. Releasing stepovers with linking normal faults are more conducive to through-going rupture than restraining stepovers with linking thrust faults. This disparity is driven by the sign of the normal stress increment: strike-slip displacement induces a negative increment (unclamping) on linking normal faults, whereas it causes a positive increment (clamping) on linking thrust faults. These dynamics, alongside significant rake rotations at fault intersections, determine whether a stepover acts as a seismogenic barrier or a nucleation site, underscoring the importance of two-way stress interactions in geometrically complex fault systems. Wesnousky (2006) evaluated these results. Releasing stepovers should be kinematically more conducive to rupture propagation than restraining stepovers; paradoxically, it is the former that are more frequently observed as rupture termination points. This apparent discrepancy is resolved by the fact that releasing stepovers outnumber restraining stepovers by a ratio of six to one within the current dataset. While restraining geometries may be intrinsically more effective at arresting individual seismic ruptures, cumulative strike-slip deformation appears to favour the creation and preservation of extensional features, while simultaneously promoting the linkage and eventual removal of restraining stepovers. In addition to the difference between releasing and restraining stepovers, another compelling topic is the critical jump distance, which is the maximum rupture jumping distance across stepovers. An early statistical analysis by Knuepfer (1989) indicated that most strike-slip earthquakes fail to jump a releasing stepover wider than 8 km and a restraining stepover wider than 5 km. Barka and Kadinsky-Cade (1988) proposed that 5 km could be a critical jump distance. Wesnousky (1988) compiled data from the USA, Turkey, and Japan, combining it with his further statistics in Wesnousky (2006), and this critical jump distance was modified to 3–4 km. Extensive research has targeted this threshold, yet it remains a subject of ongoing debate. The identified control factors include fault geometries (Harris et al., 1991; Harris and Day, 1993; Duan and Oglesby, 2006; Shaw and Dieterich, 2007; Mia et al., 2024), friction (Ryan and Oglesby, 2014), and interactions (Segall and Pollard, 1980; Yin et al., 2023), initial stresses (Lozos and Harris, 2020; Wang et al., 2020a), seismogenic depths (Bai and Ampuero, 2017), topography (Douilly, 2023), and poroelasticity (Huang et al., 2025).

Another important role of stepovers in fault systems is in fault linkage. Strike-slip faults always develop multiple stepped segments in the plan view, which are connected by stepovers, where faults interact and stress is distributed heterogeneously, forming various structures including restraining bends, releasing bends, horsetail splay, and en échelon folds (Kim and Sanderson, 2006; Mann, 2007). These structures are manifested as positive (Fig. 6d) and negative (Fig. 6b) flower structures in the cross-section view. As the strike-slip faults evolve, the stepovers accumulate damage, linking discontinuous strike-slip faults and forming a principal displacement zone (PDZ) (Sibson, 1985; Choi et al., 2016). Numerous theoretical, experimental, and field investigations employed Riedel shears to interpret the incipient evolution of strike-slip faults. The initiation of a strike-slip fault system is characterised by the development of discrete, en échelon arrangements of R-shears, typically oriented at an acute angle of 10–20° to the PDZ. This initial stage is followed by the emergence of antithetic R’-shears and the subsequent formation of P-shears, which develop symmetrically to the R-shears relative to the fault strike. As deformation progresses, Y-shears manifest near-parallel to the main fault axis. These secondary shears gradually coalesce into a complex staircase trajectory (Sylvester, 1988; Martel and Pollard, 1989; Peacock and Sanderson, 1995; Aydin and Berryman, 2010). As strike-slip faults evolve, fault segments or Riedel shears would interact with each other by forming damage zones (Segall and Pollard, 1983; Cruikshank et al., 1991; Kim et al., 2004). The linking damage zones at stepovers are considered to be more complex than others. Kim et al. (2004) described the linkage between mode II tips of strike-slip faults. For releasing stepovers, they summarised four types of structures: (1) Extension fractures (Fig. 7a). Linkage is mainly driven by $\sigma_1$-parallel extension fractures, such as joints or veins, where $\sigma_1$ is the maximum principal stress. These fractures terminate against existing fault planes or act as linking structures (Segall and Pollard, 1983; Peacock and Sanderson, 1995). (2) Pull-aparts (Fig. 7b). Increasing slip on the fault segments further increases the aperture of extension fractures, forming pull-aparts (Aydin and Nur, 1985). (3) Rotated blocks (Fig. 7c). Increasing slip drives synthetic block rotation, forming triangular apertures at intersections. These blocks are bounded by antithetic faults that initiate at acute angles but may rotate into obtuse orientations as displacement increases. The rotation magnitude is proportional to fault slip and is governed by stepping geometry and fault-normal stress (Gross et al., 1997). (4) Strike-slip duplexes or isolated lenses (Fig. 7d). Extensional apertures associated with fault-bounded blocks typically host vein mineralisation or sedimentary infill. Such duplex systems are commonly subjected to breaching, where through-going faults truncate the imbricate structures to link the primary stepping segments (Woodcock and Fischer, 1986). For restraining stepovers, they summarised three types of structures: (1) Rotated blocks (Fig. 7e). Sigmoidal fault traces within the stepover signify distributed simple shear or stress-field perturbations caused by fracture interaction. Antithetic kinematics are confirmed by secondary extension fractures at the fault tips (Ramsay, 1983; Olson and Pollard, 1991), which serve as reliable kinematic indicators for the sense of displacement within the overstep. (2) Connecting faults (Fig. 7f). Strike-slip relay ramp, veins, antithetic faults, and pressure solution seams can occur within the fault stepover (Bürgmann and Pollard, 1994; Peacock and Sanderson, 1995). (3) Strike-slip duplexes or isolated lenses (Fig. 7g), despite being less common than those in releasing stepovers. Compared to strike-slip faults, normal faults also develop linking damage zones, but focus more on the interpretation from the view of relay structures, whose formation and destruction are considered to be the most efficient way for faults to lengthen (Fossen and Rotevatn, 2016). Proximity between sub-parallel faults triggers interaction, leading to fault-tip retardation, geometric curving, and increased subsidiary fracturing (Peacock and Sanderson, 1991). This process generates steep displacement gradients that manifest as a relay ramp, elevating tip stress concentrations and driving propagation into the adjacent stress drop region; however, propagation arrests at a critical stress-drop threshold (Gupta and Scholz, 2000). This interaction results in a soft-linked relay ramp, which may evolve into a hard-linked breached relay under continued strain (Walsh and Watterson, 1991; Childs et al., 2017). Fossen and Rotevatn (2016) summarised the breaching modes of relay ramps: (1) Single-tip breaching. One fault tip arrests, while the other deflects to intersect the adjacent segment, creating a characteristic curved linkage in map view. (2) Double-tip breaching. Both faults propagate and connect, forming a lenticular relay zone. (3) Mid-ramp breaching.

Canyonlands faults favour this mode, as the strata may be more brittle. Regardless of fault regime, linking damage zones are highly complex, governed by multiple factors including fault geometry and rock properties. In this context, researchers have commenced leveraging statistical correlations to derive quantitative insights. A well-established scaling-law relationship exists between fault trace length $L_f$ and maximum displacement $d_{\max}$ (Cartwright et al., 1995; Kim and Sanderson, 2005):

$$d_{\max} = c_f L_f^{\,n} \tag{4}$$

where $n$ ranges from 0.5 to 2 (Walsh and Watterson, 1988; Marrett and Allmendinger, 1991; Cowie and Scholz, 1992; Gillespie et al., 1992; Scholz et al., 1993; Clark and Cox, 1996), and $c_f$ is an expression of fault displacement at unit length. Kim and Sanderson (2005) summarised factors controlling the ration of $d_{\max}$ and $L_f$, including (1) material property; (2) type of fault including position and direction of measurement; (3) earthquake rupture and slip/propagation history; (4) segmentation and linkage (evolution); and (5) reactivation. These $d_{\max}/L_f$ ratios are also widely used in fault growth models, including the propagating fault model and linkage model. Lathrop et al. (2022) collected normal fault $d_{\max}/L_f$ through time data, and they found that $d_{\max}/L_f$ ratios are changing throughout the evolution of faults; therefore, it is essential to exercise caution when interpreting static $d_{\max}/L_f$ data.

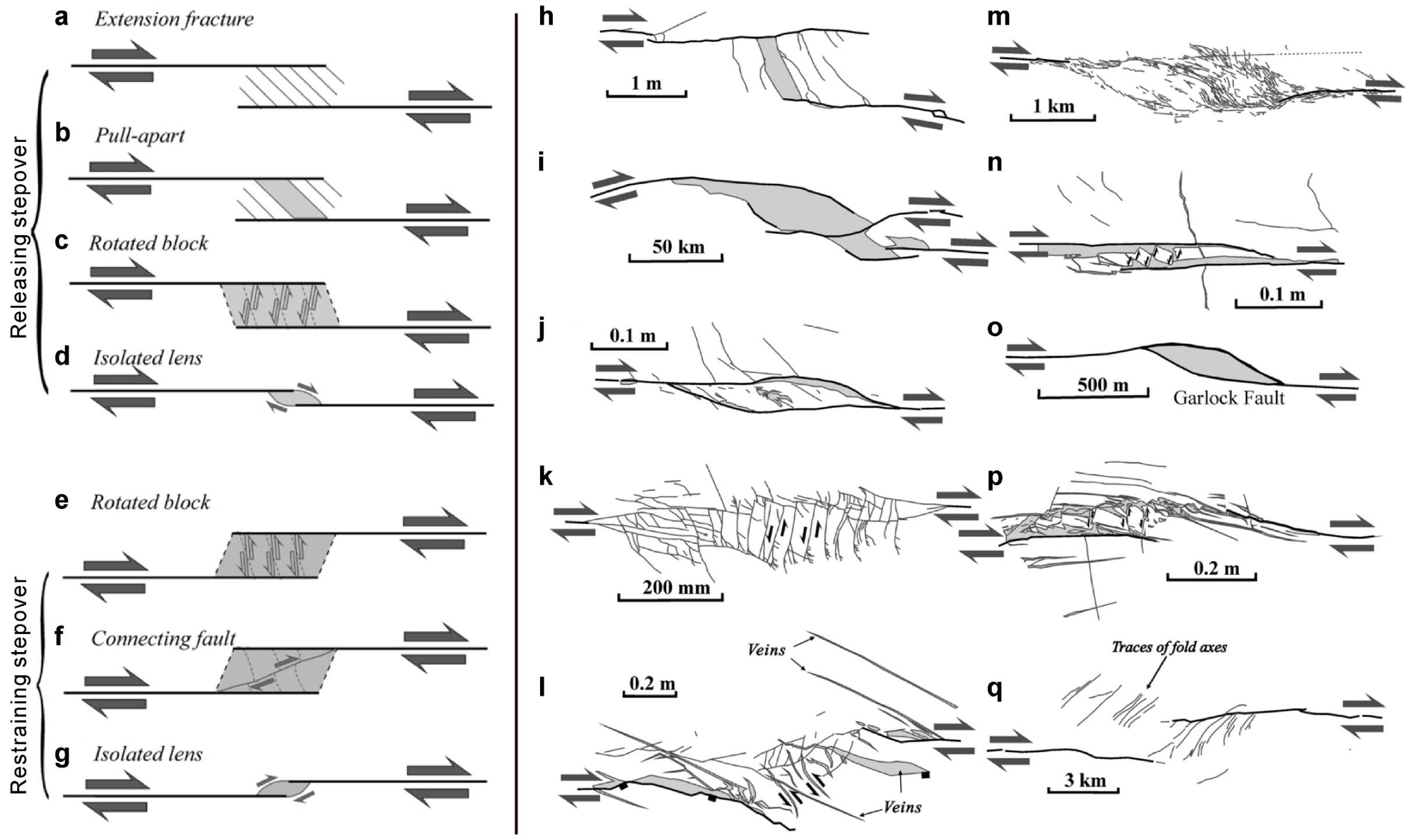

**Fig. 7. Failure modes of the stepover (rock bridge) in strike-slip faults, modified from Kim et al. (2004). (a–d)** Failure modes of releasing stepovers, including **(a)** extension fracture, **(b)** pull-apart, **(c)** rotated block, and **(d)** isolated lens. **(e–g)** Failure modes of restraining stepovers, including **(e)** rotated block, **(f)** connecting fault, and **(g)** isolated lens. **(h–q)** Examples of linking damage zones in strike-slip faults summarised by Kim et al. (2004). **(h)** Extension fractures and pull-aparts on Gozo, Maltese islands (Kim, 2000). **(i)** Extension fractures and strike-slip duplexes at Dasht-e Bayaz, Iran (Tchalenko and Ambraseys,

1970). **(j)** The Vienna Basin, Austria and Czechoslovakia, which is a pull-apart basin (Royden, 1985). **(k)** Block rotation at Kilve, Somerset (Kelly, 2001). **(l)** Strike-slip duplexes or isolated lenses at Rame Head, Cornwall, UK (Kim et al., 2001). **(m)** Isolated lenses from Cantil Valley, California, USA (Aydin and Nur, 1985). **(n)** Block rotation from Gozo, Maltese islands (Kim, 2000). **(o)** Block rotation from Kilve, Somerset, UK (Kelly, 2001). **(p)** Antithetic faults and connecting faults from East Quantoxhead, Somerset, UK (Peacock and Sanderson, 1995). **(q)** Potential linkage in the Coyote Creek fault zone, California, USA (Sharp and Clark, 1972). Thick line is the major fault, thin line is the minor fault, shading is the vein in small scale examples and basin in large-scale examples.

Rock bridges also manifest as cohesion-healed patches in fault systems due to fault healing through processes such as pressure solution during interseismic periods (Muhuri et al., 2003; Tenthorey et al., 2003; Tenthorey and Cox, 2006; Pei et al., 2019; Luo et al., 2022). Tenthorey et al. (2003) documented that strength recovery in experimental faults, driven by cohesion-healed patches, reached 75% of the original intact rock strength. Recent work has already used rock bridges as laboratory analogues for the cohesion-healed patches to investigate fault heterogeneity (Luo et al., 2022; Luo et al., 2026).

Regardless of the manifestation in fault systems, rock-bridge fracturing is a highly complex process governed by various factors, such as geometry and boundary conditions. Current approaches primarily rely on observational classification or statistical $d_{\max}$–$L_f$ relationships, and laboratory simulations remain highly simplified (see Section 7 for more examples). A unified mechanical framework to analyse rock bridges is yet to be established.

### 4.4. Mechanistic insights into the importance of rock bridges

The importance of rock bridges arises primarily from the significant contrast between the mechanical properties of the rock matrix and those of discontinuities. Müller (1966) noted that the deformation resistance of rock bridges is mobilised at significantly lower displacement thresholds than those required to activate discontinuity resistance. Von Thun and Tarbox (1971) further validated that a very few rock bridges would cause a substantial increase in the overall shear modulus of rock masses.

In addition to the significant contrast in deformation properties between rock bridges and discontinuities, strength contrast may be more intuitive. ISRM (1978) defined discontinuities as those having zero or low tensile strength. Despite this definition neglecting some stronger incipient traces (Hencher and Richards, 2015), most discontinuities present much lower tensile strength than rock bridges. An analysis of a natural rock bridge under direct shear conducted by Hencher (1984) quantified the strength contribution provided by the intact rock bridge. Direct shear testing revealed that these areas provided a cohesive contribution of 750 kPa. Shang et al. (2016) conducted uniaxial tensile tests on natural rock samples containing visible, incipient discontinuities including joints, bedding, and mineral veins. Results reveal that the tensile strength of incipient discontinuities is mainly attributed to rock bridges. Further analysis indicates that, in siltstone, the uniaxial tensile strength *T* of incipient bedding planes was found to range from 32% to 88% of that of the parent rock (Shang et al., 2018). Comparatively, joints within the same lithology exhibited *T* values between 23% and 70% of the parent material, with rock bridges accounting for 23% to 70% of the total discontinuity surface. In addition, the shear strength of the rock matrix is typically two or more orders of magnitude greater than that of discontinuities (Kemeny, 2004; Barton, 2011). Moreover, a rock bridge is always treated as an increase in cohesion that can be either represented by an implicit cohesion (Jaeger, 1971; Eberhardt et al., 2004a; Karami et al., 2007), explicit cohesion (Jennings, 1970; Stead et al., 2006; Sturzenegger and Stead, 2012; Alghalandis et al., 2017), or both (Pariseau et al., 2008).

## 5. Measurement

### 5.1. Geometric descriptors of rock bridges

Rock-bridge definition indicates that discontinuity arrangements determine rock-bridge configurations. Early conceptualisations of rock bridges were restricted to a model with collinear discontinuities, namely, in-plane rock bridges (Fig. 8a). As such, the concept of discontinuity persistence *K* can be applied to quantify rock bridges. However, natural discontinuities typically appear in sets that are more or less parallel and regularly spaced in natural rock masses. For discontinuities deviated from collinear arrangements, namely, out-of-plane discontinuities, the concept of *K* does not work. Early rock-bridge research described the bridging geometry using descriptors of discontinuities (Fig. 8b), such as discontinuity continuity *C*, discontinuity spacing *h*, and discontinuity inclination angle *β*. Nevertheless, such descriptors may be more complex when multiple discontinuities exist or overlapping discontinuities present. For simplicity, Einstein et al. (1983) proposed the concept of a ligament (Fig. 8c), defined as the line connecting discontinuity tips, and bridging angle *α*, defined as $\sin^{-1}(h/L)$, where *L* is the ligament length. Since then, *L* and *α* have been widely used in rock-bridge research, despite certain exceptions.

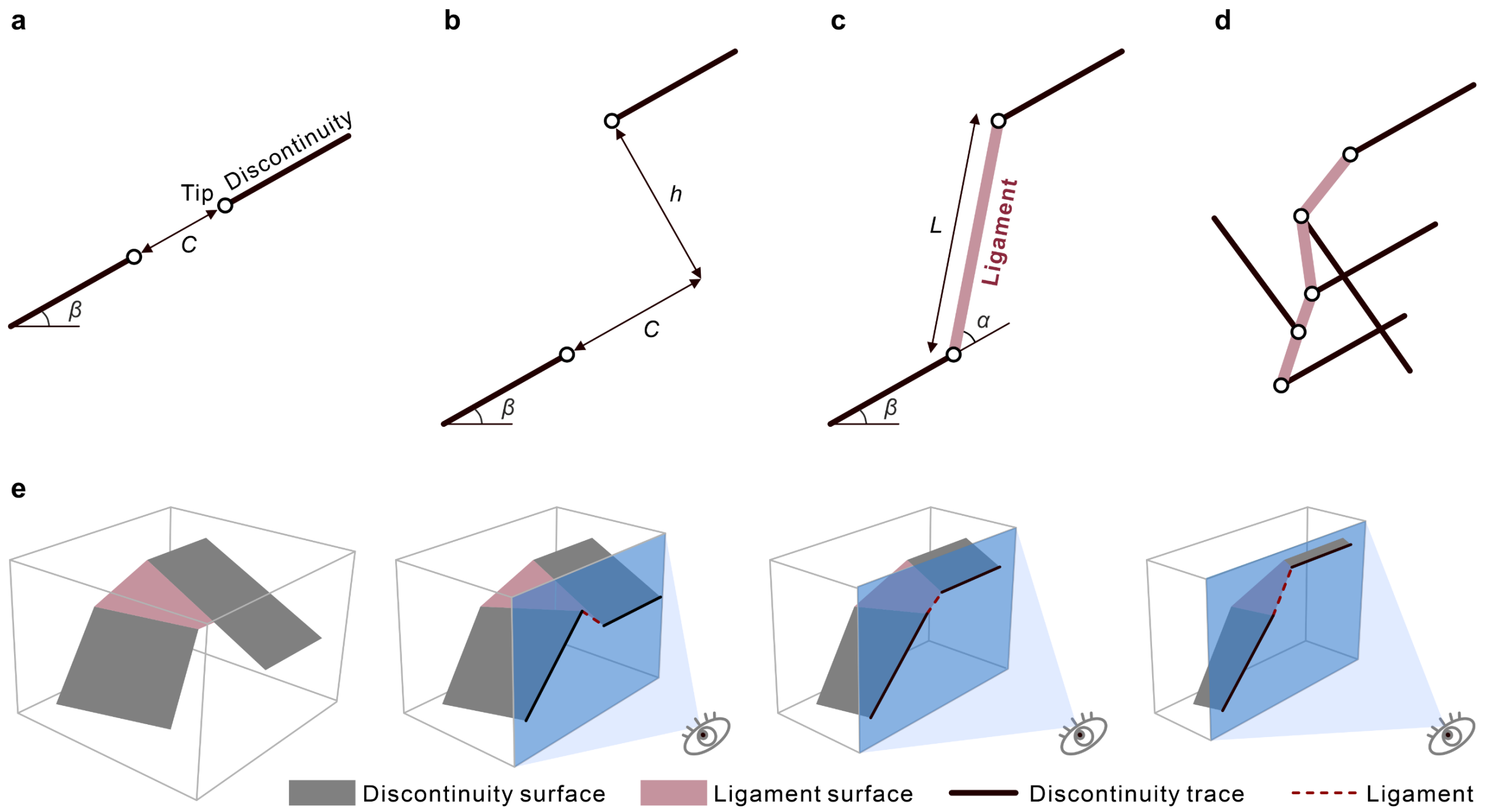

**Fig. 8. Geometric descriptors of rock bridges. (a)** A rock bridge defined by collinear discontinuities, where $\beta$ is the discontinuity inclination angle, and $C$ is the continuity. **(b)** A rock bridge defined by discontinuities deviate from collinearity, where $h$ is the discontinuity spacing. **(c)** Rock bridge characterisation metrics, where $L$ is the ligament length, and $\alpha$ is the bridging angle defined as $\sin^{-1}(h/L)$, and ligament is the line connecting discontinuity tips. **(d)** Rock bridges defined by irregular discontinuities. **(e)** An irregular rock bridge in 3D. As the mapping window separates by minor offsets, rock bridge traces will be different.

Natural discontinuities can be more complicated, as well as natural rock bridges. As shown in Fig. 8d, although rock-bridge descriptors can identify ligaments, the interpretation of rock bridges would be more complex. When extended to 3D, the geometry of rock bridges would be more difficult to describe. As shown in Fig. 8e, discontinuities can have different orientations in 3D, resulting in irregular configurations of rock bridges in 3D. In this case, the ligament can be extended from a linear to a surface representation, though its mapping would be difficult and is detailed in subsequent sections.

### 5.2. Rock-bridge estimation

The measurement of rock bridges originates with the estimation of discontinuity persistence $K$ by Stini and Müller (Müller, 1962), which the original text identifies as "degree of separation", as defined by Eqs. (1)–(3). Since the conceptualisation of a rock bridge, index $(1 - K)$ has been the common measurement for rock bridges, namely, rock-bridge percent.

#### 5.2.1. Forensic investigation

The inherent difficulty in mapping the full spatial distribution of discontinuities within a rock mass complicates the accurate estimation of rock-bridge percent. In most cases, the complete morphology of discontinuities and rock bridges can only be precisely observed once a rock mass fails and the rupture surface is exposed. The approach of measuring rock bridges through analysing the characteristics of the rupture surface after failure is always referred to as the forensic investigation (de Vilder et al., 2017; Shang et al., 2017). While retrospective, this measurement is essential for elucidating the progressive failure of a rock mass. As exemplified by landslide research, most estimations of rock-bridge percent are based on forensic investigations. Supplementary to the data collected by Tuckey and Stead (2016), further estimations of rock-bridge percent in landslides are presented in Table 1.

**Table 1. Estimation of rock-bridge percent in landslides, supplementary to the data by Tuckey and Stead (2016).**

| Location | Measurement method | Rock bridge percent (%) | Reference |
|---|---|---|---|
| Highway roadcuts, Idaho, USA | Failure mapping; length % | 0 to 3 | (Ristau, 1994) |
| Rockfall scars, French Sub-Alps | Failure mapping; area % | 0.2 to 5 | (Frayssines and Hantz, 2006) |
| Failed overhang road half-tunnel, Italy | Failure mapping; area % | 26 | (Paronuzzi and Serafini, 2009) |
| Åknes slide, Norway | Failure mapping; length % | 1 to 3 | (Grøneng et al., 2009) |
| Gypsum mine, Arizona, USA | Failure mapping; length % | 16 to 36 | (LeBaron, 2011) |
| Diavik A154 Pit, Northwest Territory, Canada | DFN modelling study; length % | >5 | (Karami et al., 2007) |
| | | 25 | (Moffitt et al., 2007) |
| | | 3 to 45 | (Elmo et al., 2011) |
| | | 5 to 10 | (Sampaleanu, 2017) |
| Natural slope failure, Western Alps, France | Failure mapping; area % + numerical simulation | 7.3 | (Lévy et al., 2010) |
| Randa rockslide, Switzerland | Length % for specific discontinuity sets + numerical simulation | 8 to 45 | (Gischig et al., 2011) |
| Half Dome rockfall, California, USA | Failure mapping; area % | 26 | (Stock et al., 2011) |
| Palliser slide, Alberta, Canada | Failure mapping; length % | 2 to 3 | (Sturzenegger and Stead, 2012) |
| Mont Blanc rockfalls, Drus Peak, France | Failure mapping; area % | 3 | (Matasci et al., 2015) |
| Néron rockfall, Frace | Failure mapping; area % | 5 | (Bonilla-Sierra et al., 2015) |
| Rockfall scars, North Yorkshire coastal cliffs, UK | Failure mapping; area % | 0 to 97.6 | (de Vilder et al., 2017) |

However, the concept of rock-bridge percent is constrained to in-plane rock bridges. In other cases, despite the current lack of a systematic metric, measuring rock-bridge profiles of specific cases remains of critical importance. Rock-bridge profiles can be distinguished from pre-existing discontinuities by the lack of staining or weathering and by the conchoidal or irregular surface texture of the fracture surface (Tuckey and Stead, 2016), which is a common methodology to measure rock bridges. Frayssines and Hantz (2006) suggested that rock-bridge failure is geologically indicated by the presence of fresh intact rock (Fig. 9a–c); specifically, light beige to rosy hues for Urgonian limestone and dark grey for Thitonian limestone. These nascent fracture surfaces contrast sharply with the established scar, which is typically obscured by a calcite crust ranging from white to orange depending on iron oxide concentrations. Although these fresh surfaces often constitute a negligible portion of the overall scar area, they represent critical failure zones. Such failure manifests across translational sliding surfaces (shear mode), rear tension cracks (tensile mode), or small basal surfaces, which may function as either sliding or opening planes during toppling events. Using a similar method, in the scar of the 1991 Cellina rockfall, Paronuzzi and Serafini (2009) measured a rock bridge on the contact surface by the distinct grey-to-pale-brown hue of the freshly failed intact rock, which contrasts sharply with the yellowish, weathered joint (Fig. 9d). Furthermore, the rock bridge is characterised by a markedly concave morphology, distinguishing it from the characteristically flat profile of the weathered fracture joint. Paronuzzi et al. (2015; 2016a) also identified a rock bridge with a size of ~15 cm × 30 cm in the scar of the 1983 Rosandra wedge collapse (Fig. 9e,f), as the light grey-coloured area (fresh colour of the intact rock) of the rock bridge was clearly different from the surrounding yellowish-brown weathered discontinuity surfaces. Other landslide cases containing rock bridges that are identified using similar methodology include the 2004 Rosandra rockslide (Paronuzzi and Bolla, 2022), 2007 Chamousset limestone column collapse (Lévy et al., 2010), 2008 Half Dome rockfall (Stock et al., 2011), 2011 Néron rockfall (Bonilla-Sierra et al., 2015), 2014 San Leo landslide (Donati et al., 2019), 2019 Hüttschlag rockfall (Gerstner et al., 2023), Palliser Rockslide (Sturzenegger and Stead, 2012), multiple rockfalls on the North Yorkshire coast (de Vilder et al., 2017), Diavik A154S rockfall (Sampaleanu, 2017), Weeping Wall rockfall (Donati et al., 2023), etc.

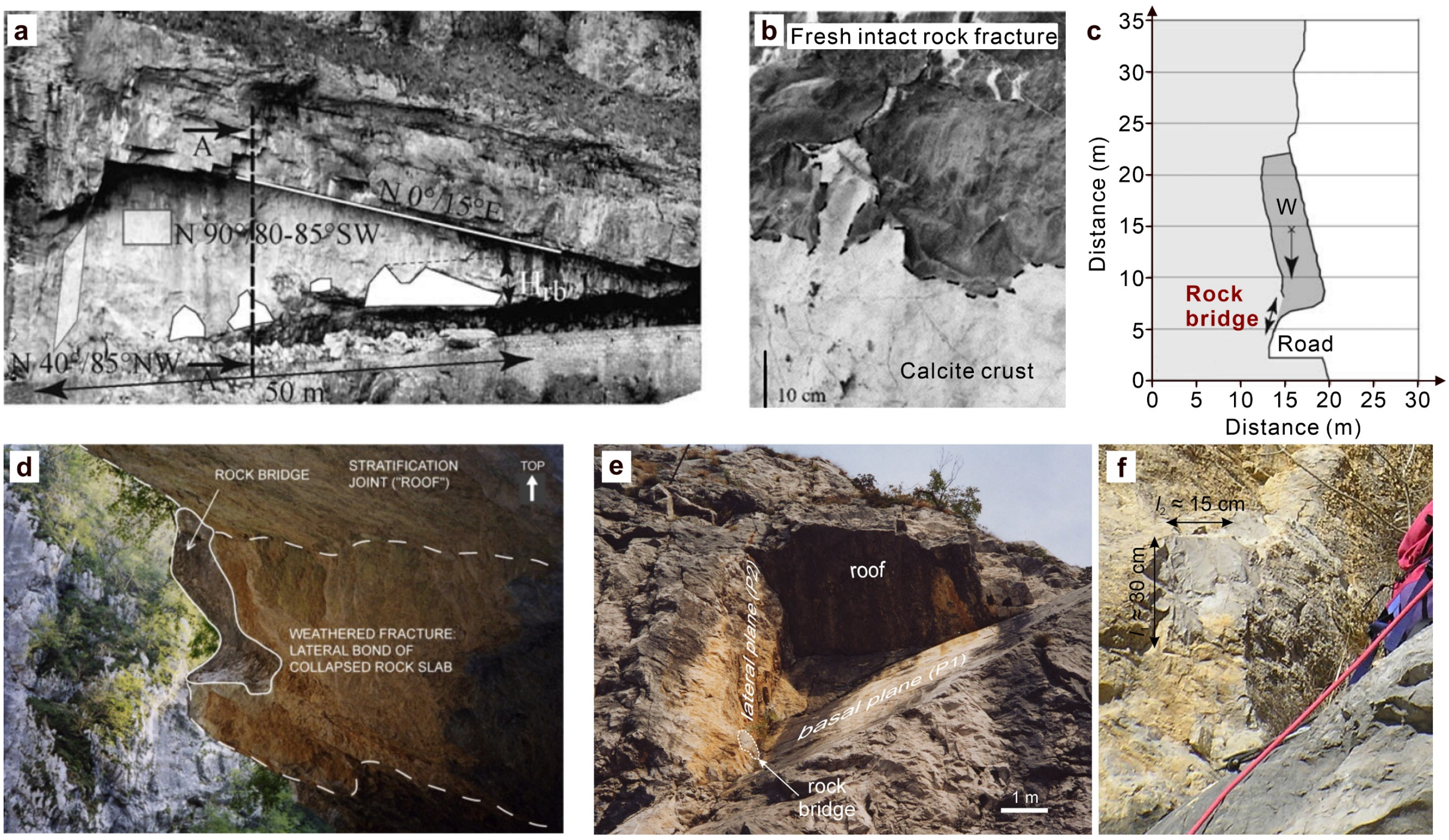

**Fig. 9. Rock-bridge profiles in landslides. (a)** Rock bridges (white parts) in the Ranc rockfall (Frayssines and Hantz, 2009). **(b)** Boundary between the rock bridge and failure surface. **(c)** Cross section of the cliff before failure. **(d)** A rock bridge in the 1991 Cellina rockfall (Paronuzzi and Serafini, 2009). **(e)** Rock-bridge profile in the 1983 Rosandra wedge collapse (Paronuzzi et al., 2015; 2016a). **(f)** Close-up view of the rock bridge profile in the 1983 Rosandra wedge collapse.

For the forensic investigation of rock bridges in other domains, here are also some examples. In a direct shear test on a rock core, Hencher (1984) measured a rock bridge. The discontinuity surface is predominantly red-brown and coated with iron oxides, yet close-up observations reveal grey intact rock with minor chemical decomposition, separated by discoloured boundaries where the rock bridge is progressively being consumed by weathering processes (Fig. 10a). Shang et al. (2017) injected an expansive chemical splitting agent into incipient discontinuities of metre-scale natural rock masses (Fig. 10b). Upon expansion, the agent generated sufficient internal pressure to induce tensile failure along these latent traces, effectively transforming them into fully dilated, open joints. Consequently, rock bridges are exposed. Observations from this experimental series indicate that freshly ruptured surfaces across intact rock bridges exhibit a markedly higher rugosity compared to the smoother planes of geologically developed, albeit incipient, discontinuities. Fractographic signatures, most notably hackle and rib markings, are diagnostic of the recently failed rock bridges, whereas the activated surfaces of incipient features remain comparatively planar. Furthermore, Schmidt hammer rebound values are consistently elevated within the rock-bridge zones, likely reflecting a higher degree of geomechanical competence and a relative lack of chemical or physical weathering.

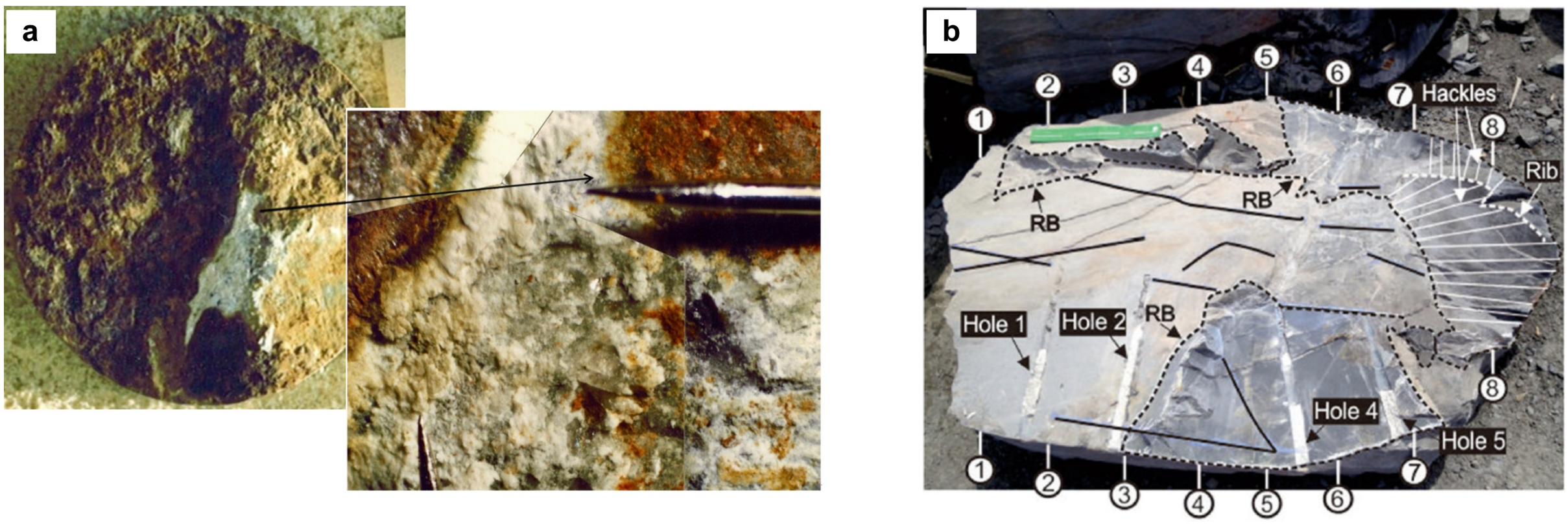

**Fig. 10. Rock-bridge profiles in rock masses. (a)** A rock bridge (white section) in a rock core subjected to direct shear (Hencher, 1984). **(b)** Rock-bridge profiles in incipient discontinuities subjected to direct tension (Shang et al., 2017). RB denotes rock bridge.

#### 5.2.2. Proactive investigation

Although the precise rock-bridge profiles can only be determined once rupture surfaces are exposed in most cases, proactive measurement of rock bridges is of paramount importance, as it directly informs stability analysis. The proactive measurement of rock bridges can be approximately estimated through various geological investigations, including outcrop mapping, photogrammetry, borehole drilling, TLS, ALS, interferometric synthetic aperture radar (InSAR), infrared thermal imaging, ground-penetrating radar, and ultrasonic testing. Outcrop mapping offers an intuitive means of measuring rock bridges. Therefore, numerous researchers mapped rock-bridge profiles in outcrops, such as the outcrop of Mount Edith Cavell (Fig. 11a) measured by Sturzenegger and Stead (2009), the outcrop of Decker Mountain, BC Skier (Fig. 11b) measured by Sampaleanu (2017), and the outcrop of US Highway 93 near Wickenburg, Arizona, USA, measured by Monte (2004). Notably, Monte (2004) formally incorporated the outcrop-measured rock-bridge percent into rock mass classification systems.

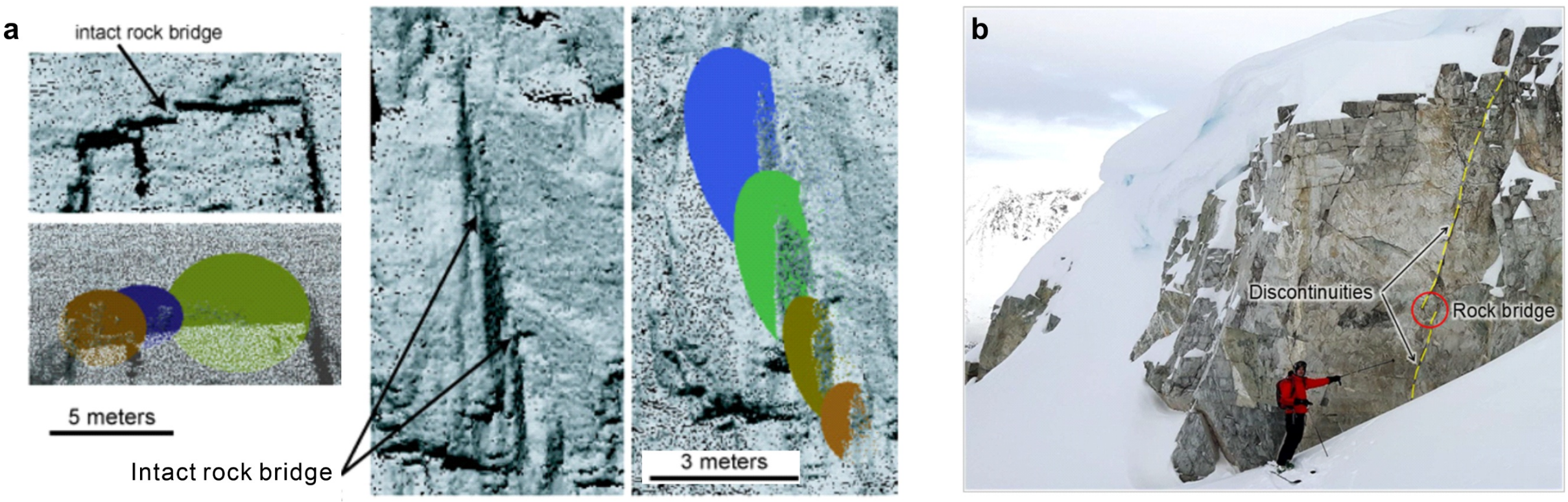

**Fig. 11. Rock-bridge profiles measured in outcrops. (a)** Rock-bridge profiles in the outcrop of Mount Edith Cavell (Sturzenegger and Stead, 2009). **(b)** Rock-bridge profile in the outcrop of Decker Mountain, BC Skier (Sampaleanu, 2017).

Nevertheless, outcrop-measured rock bridges entail uncertainty, as subsurface structures are difficult to ascertain precisely. Generally, geological mapping represents discontinuity length as the diameter of a disc or the edge of a polygon inscribing the mapped feature. Regardless of the assumed geometry, the mapped discontinuity length remains an explicit 2D measure of the trace intersecting a geological exposure. Consequently, this linear trace does not necessarily correspond directly to the underlying discontinuity, as evidenced by the natural discontinuity and rock-bridge profiles by Shang et al. (2017) in Fig. 10b. Outcrop measurement of rock bridges faces the same challenges. Figure 8e demonstrates the inherent challenges of measuring a rock bridge within a 2D window map. The spatial positioning and orientation of the sampling window exert a profound influence on the resulting 2D characterisation; notably, three parallel mapping windows, separated only by minor offsets, yield markedly disparate interpretations of the rock-bridge configuration.

In the Åknes potential rockslide, based on the geological mapping, geophysical surveys, and drilling of boreholes, Grøneng et al. (2009) interpreted the internal structure, especially the rock-bridge profile with a percent of 1–3% (Fig. 12a). Although Grøneng et al. (2009) noted that the exact configuration of the rock bridge is not finally defined, it represents a realistic alternative. Geophysical surveys are considered promising to measure rock bridges (Alcaino-Olivares, 2024). For instance, in the Zhengjiadagou potential landslide, Wang et al. (2015)

employed field mapping, geophysics, and multi-depth displacement monitoring (surface and borehole) to determine a comprehensive deformation zoning of the slope in both the terrestrial and subsurface domains. Especially, they identified several locked segments (rock bridges) by transient electromagnetic method (Fig. 12b). Notwithstanding the potential uncertainties, such as water and lithology, this approach represents a promising avenue for measuring rock bridges.

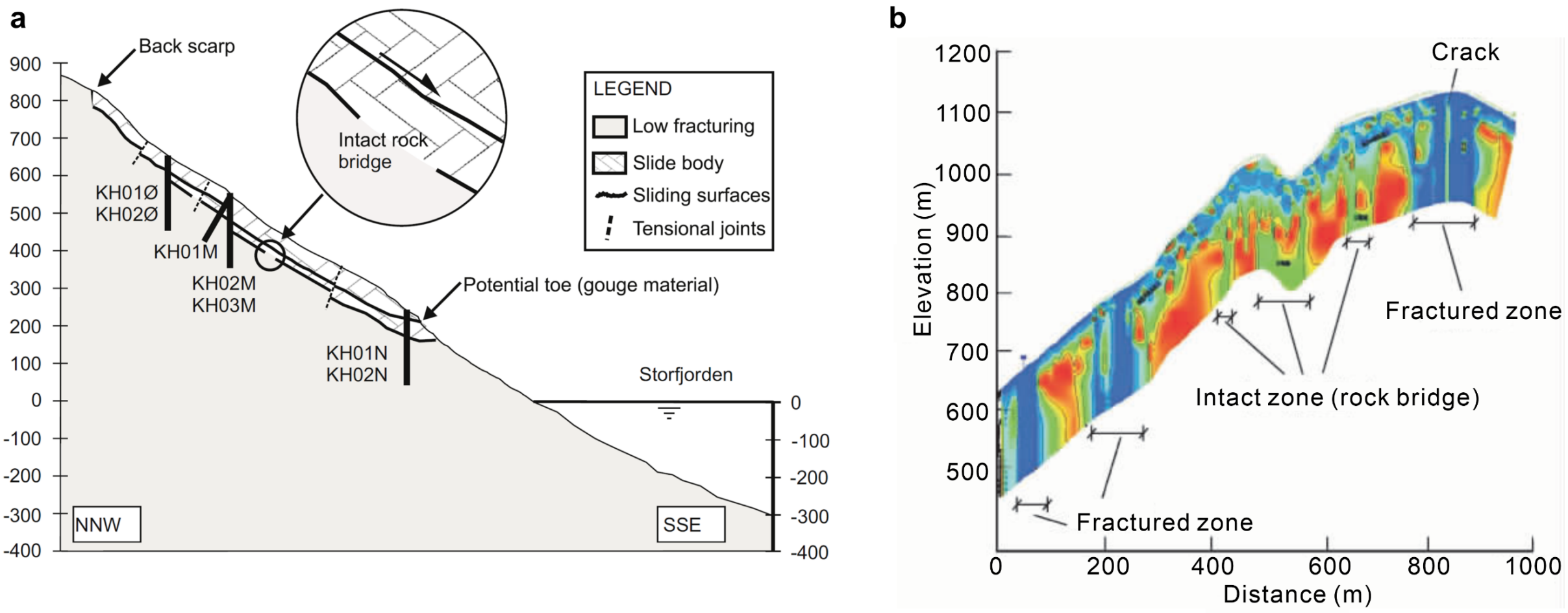

**Fig. 12. Proactive investigation of rock-bridge profiles in landslides. (a)** Rock-bridge profile in the Åknes potential rockslide (Grøneng et al., 2009). **(b)** Rock-bridge profiles in the Zhengjiadagou potential landslide by transient electromagnetic method, modified from Wang et al. (2015).

Recent advancements have employed remote sensing to measure rock bridges prior to failure. A representative case is the application of infrared thermal imaging by Guerin et al. (2019) to measure rock bridges. Based on a potential rockslide in an exfoliating cliff of El Capitan within Yosemite Valley, California, USA, namely, the Boot Flake, they conducted infrared thermal panoramic mapping. Advective cooling from air circulation along the detachment interface results in a predominantly cold thermal signature across the exposed sheet. However, persistent warm thermal anomalies are detected near the centre of the Boot Flake, spatially coinciding with the rock bridge identified during climbing investigations (Fig. 13). The thermal contrast remains discernible even during nocturnal cooling phases, allowing for the differentiation between the coupled rock bridge (thermal highs) and the detached surface (thermal lows). Integrating these geometries into a numerical thermal model yields temperature gradients and anomalous patterns that align closely with infrared panoramic mapping, thereby validating the interpreted rock-bridge profile. As noted by Elmo et al. (2021a), although very promising, this approach appears constrained to geometries characterised by cliff-parallel exfoliation fractures and partially decoupled rock sheets.

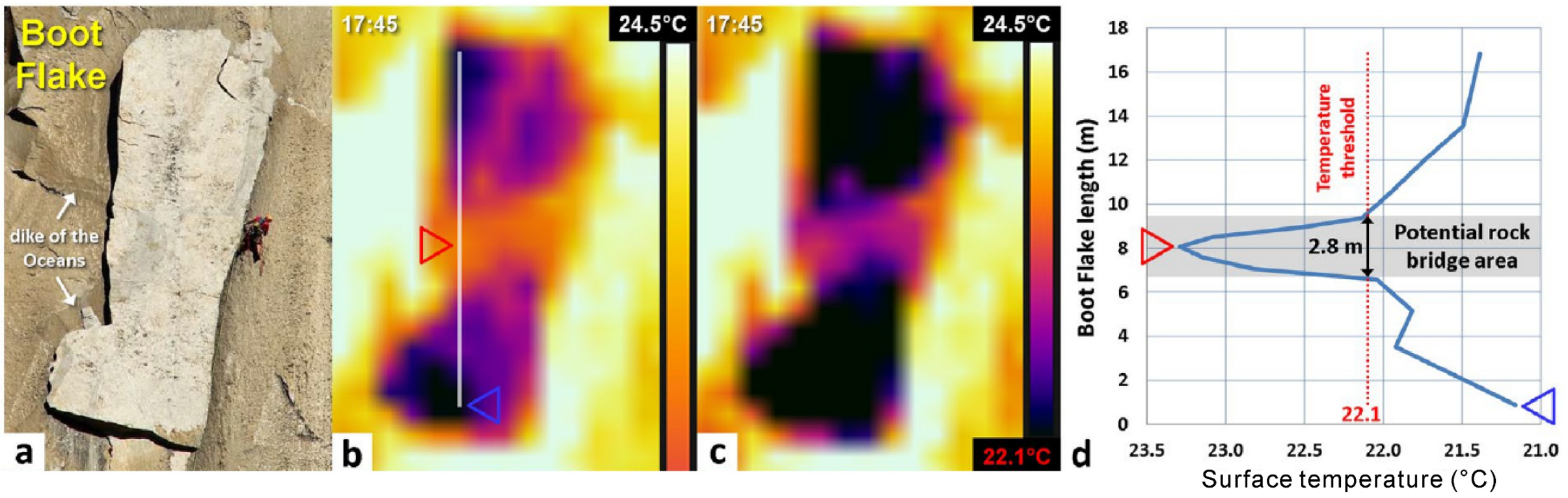

**Fig. 13. Rock-bridge profiles in the Boot Flake potential rockslide (Guerin et al., 2019).**

There are also fundamental studies aimed at measuring rock bridges. For example, Tang et al. (2024) tried to identify rock-bridge profiles by taking $\sigma_1$ as the index of k-means clustering. This approach essentially maps zones of $\sigma_1$ concentration and equates them with rock bridges. Numerous field and experimental investigations have demonstrated that rock bridges are subjected to complex stress states; therefore, relying solely on $\sigma_1$ concentration may introduce uncertainty when measuring rock bridges.

Despite the ongoing development of proactive investigations of rock bridges, existing studies have shown that establishing a correlation between rock-bridge configurations and mechanical behaviour is a prerequisite for further reliable measurement.

### 5.2.3. Probabilistic and statistical methods

Given the abovementioned difficulty in measuring rock bridges, Einstein et al. (1983) developed probabilistic models employing a method of slices alongside Monte Carlo simulations to quantify slope failure probability as a function of joint geometry and the relative resistance of rock bridges and discontinuities. Similarly, stochastic modelling of discontinuity geometries has been utilised to evaluate the kinematic feasibility and kinetic stability of rock slopes, specifically focusing on the clustering and connectivity of natural joint systems (Dershowitz and Einstein, 1988). These studies underscore the critical role of discontinuity geometry, persistence, and rock bridges, alongside their statistical distributions, in the formation of potential failure surfaces.

Some researchers use the discrete fracture network (DFN) framework to measure rock bridges (Elmo et al., 2007; Karami et al., 2007; Moffitt et al., 2007), which is an optimal numerical method for synthesising realistic fracture networks and rock bridges by integrating data from both digital and conventional mapping. Tuckey et al. (2012) and Elmo et al. (2018) further developed this methodology, and the index to characterise rock bridges is named $RB_{ij}$ (Table 2), where the subscripts i and j denote the dimensions of sample and measurement, respectively.

In this way, $RB_{22}$ can be defined as unity minus the proportion of the discontinuity surface area represented in Eq. (1), and $RB_{11}$ can be defined as unity minus the proportion of the discontinuity line represented in Eq. (2). Tuckey et al. (2012) and Sampaleanu (2017) already employed $RB_{10}$ and $RB_{21}$ to quantify rock bridges. Elmo et al. (2018) noted that because linear and areal rock bridge densities are anisotropic properties, any persistence measurement derived from these values will inevitably exhibit orientation-induced sampling bias.

**Table 2. $RB_{ij}$ system to quantify rock bridges by Elmo et al. (2018).**

| | | Dimension of feature | | | |
|---|---|---|---|---|---|
| | | 0 | 1 | 2 | 3 |
| Dimension of sampling region | 0 | **$RB_{00}$**<br>Number of rock bridges | | | |
| | 1 | **$RB_{10}$**<br>Linear rock-bridge intensity, defined as the number of rock bridges per unit length of equivalent discontinuity | **$RB_{11}$**<br>Linear rock-bridge ratio, defined as the rock-bridge length per unit length of equivalent discontinuity | | |
| | 2 | **$RB_{20}$**<br>Areal rock-bridge density, defined as the number of rock bridges per unit area of equivalent discontinuity | **$RB_{21}$**<br>Areal rock-bridge intensity, defined as the rock-bridge length per unit area of equivalent discontinuity | **$RB_{22}$**<br>Areal rock-bridge ratio, defined as the rock-bridge area per unit area of equivalent discontinuity | |
| | 3 | **$RB_{30}$**<br>Volumetric rock-bridge density, defined as the number of rock bridges per unit volume of rock mass | | **$RB_{32}$**<br>Volumetric rock-bridge intensity, defined as the rock-bridge area per unit volume of rock mass | **$RB_{33}$**<br>Volumetric rock-bridge ratio, defined as the rock-bridges volume per unit volume of rock mass |

The $RB_{ij}$ system can be compatible with rock bridges defined by discontinuities deviated from collinear arrangement by introducing the concept of equivalent discontinuities. See Elmo et al. (2018) for more information. According to this system, most forensic and proactive investigations are $RB_{00}$, $RB_{10}$, $RB_{11}$, $RB_{20}$, $RB_{21}$, $RB_{22}$, and $RB_{30}$, while $RB_{32}$ and $RB_{33}$ are conceptual and elusive to characterise morphologically. Therefore, describing a rock bridge in 3D as a ligament surface may be more feasible to measure, as illustrated in Section 5.1. Nevertheless, this does not imply that identifying the possible rock bridges functioning as blocks is unimportant; for instance, Fadakar Alghalandis and Elmo (2018) investigated the degree of interlocking from rock bridges and concluded that, as the degrees of jointing and interlocking increase, the number of available rock bridges decreases, which is important for evaluating rock masses; however, such research pertains to the mechanical rather than the geometric description of rock bridges.

Furthermore, Elmo (2023) proposed the Bologna interpretation of rock bridges. The concept is that “prior to failure it is not a matter of describing rock bridges as physical deterministic entities. There are only potential rock bridges which exist everywhere at once. Only by looking at a rock mass as it fails we can summon the position of a rock bridge into existence, and therefore measure the rock bridge potential of a rock mass." In this interpretation, rock bridge potential represents the contribution of rock bridges to the rock mass strength. This index evaluates the extent to which rock mass failure is governed by stress, structure, or their hybrid. Its determination requires four parameters: intact rock strength, loading conditions (magnitude and direction), rock mass connectivity, and rock mass interlocking, which can be quantified by the network connectivity index (NCI) rock mass classification system proposed by Elmo et al. (2021b). See Section 5.3 for more details.

#### 5.2.4. Other measurement indices

There are also some other measurement indices for rock bridges. For example, Park (2005) proposed a probabilistic analysis by treating *K* as a random variable, implemented by comparing the individual joint length with the maximum sliding dimension in each interval on the premise that the discontinuity on the failure plane was fully persistent. Wasantha et al. (2014) proposed an index named the degree of persistence (DoP) by considering the impact of discontinuous joint tips on stress distribution characteristics:

$$\mathrm{DoP} = \frac{\sum_i l_{S_i}}{n_d \times (\sum_i l_{S_i} + \sum_i R_i)} \tag{5}$$

where $n_d$ is the number of discontinuity tips. However, these indices remain confined to the concept of *K* and have not been widely used.

### 5.3. Rock bridge in rock mass classification systems

Discontinuities have been comprehensively integrated into rock mass classification systems, such as the rock mass rating (RMR) (Bieniawski, 1973), Q (Barton et al., 1974), and geological strength index (GSI) (Hoek and Kaiser, 1995). In contrast, rock bridges are relatively less considered. Kim et al. (2006) noted that rock bridges are involved in the RMR system, but their weighted strength is underestimated; the Q system gives insufficient consideration to rock bridges in practice; and the GSI descriptively refers to rock bridges as block interlocking. Several researchers have made positive contributions to refer to rock bridges in rock mass classification systems. Laubscher (1990) proposed a measurement using the average fracture frequency rating (FFR) in the mining rock-mass rating (MRMR) classification system. The rock-bridge percent $P_{rb}$ was determined by:

$$\mathrm{P_{rb}} = \frac{\mathrm{FFR} \times \mathrm{a}}{40} \tag{6}$$

where $a$ is the constant that can be determined empirically (Dempers et al., 2011). This measurement is intended to neglect the fundamental role of discontinuity persistence (Elmo et al., 2018). Kemeny et al. (2003) proposed the digital rock mass rating (DRMR), where the rock-bridge percent serves as one of the seven parameters to evaluate rock masses, which is quantified using the concept of discontinuity persistence expressed by Eqs. (1)–(3). Xu et al. (2026) further improved the DRMR as 3D DRMR, where the rock-bridge percent is replaced with the discontinuity connectivity $C_n$. They employed network topology to assess $C_n$:

$$C_n = \frac{4(N_X + N_Y)}{N_I + N_Y} \tag{7}$$

where $N_X$ is the number of the X-nodes, $N_Y$ is the number of the Y-nodes, and $N_I$ is the number of the I-nodes. The predominance of X- and Y-nodes indicates a high degree of structural connectivity; consequently, the geomechanical integrity of the rock mass is frequently compromised. This topological arrangement facilitates the formation of through-going failure planes, significantly reducing the overall shear resistance of a rock mass.

Elmo et al. (2021b) proposed a network connectivity index (NCI) based on a combination of discontinuity intensity, discontinuity density, and discontinuity intersection density, expressed by:

$$\mathrm{NCI} = \frac{\mathrm{P}_{21}}{\mathrm{P}_{20}} \mathrm{I}_{20} \tag{8}$$

where $\mathrm{P}_{21}$, $\mathrm{P}_{20}$ and $\mathrm{I}_{20}$ are the areal discontinuity intensity, areal discontinuity density, and areal discontinuity intersection density, respectively. Elmo et al. (2021a) introduced the concept of $\mathrm{NCI}_\mathrm{d}$ by considering the impact of stress-driven fractures, defined by:

$$\mathrm{NCI}_\mathrm{d} = \mathrm{NCI} + \mathrm{NCI}_\mathrm{rb} \tag{9}$$

where $\mathrm{NCI}_\mathrm{rb}$ was the calculated NCI considering only the induced fractures. The ratio between $\mathrm{NCI}_\mathrm{rb}$ and $\mathrm{NCI}_\mathrm{d}$ was defined as the rock bridge potential (RBP):

$$\mathrm{RBP} = \frac{\mathrm{NCI}_\mathrm{rb}}{\mathrm{NCI}_\mathrm{d}} = \frac{\mathrm{NCI}_\mathrm{rb}}{\mathrm{NCI} + \mathrm{NCI}_\mathrm{rb}} \tag{10}$$

In this way, the higher RBP means the higher contribution of rock bridges to the overall rock mass behaviour.

### 5.4. Measurement impasse

Although rock bridges possess a dedicated set of geometric descriptors and can be extended into 3D, the precise measurement of their configurations remains problematic. This is primarily attributable to two factors: (1) the inherent invisibility of rock masses, as current technology is incapable of accurately mapping the internal structures of large-scale rock masses. Consequently, most researchers rely on limited outcrop information to measure rock bridges, which can result in significant deviations from actual subsurface conditions (Fig. 8e). (2)

Descriptive challenges of the morphology itself. Although rock bridges may function as discrete blocks, the boundaries of such volumes are difficult to delineate, despite some conceptual indices like $RB_{33}$. As a result, employing the definition of rock bridges in Section 5.1 to measure rock bridges as ligament in 2D and ligament surfaces in 3D can be more feasible.

Most rock-bridge measurements are forensic, conducted only after instability occurs and the failure surface is exposed, where rock bridges are measured based on morphological and chromatic information on failure surfaces. Some measurements are proactive, utilising geophysical or geological investigations prior to failure, though they entail inherent uncertainty. Regardless of the method, measuring rock bridges simply as a percent is fundamentally limited. Elmo (2023) summarised six such limitations: (1) the restriction to coplanar configurations (Einstein et al., 1983); (2) the assumption that rock-bridge strength is independent of position; (3) the assumption that strength is independent of scale (Bu et al., 2022); (4) the premise that failure is amenable to limit equilibrium modelling; (5) the assumption that only specific types of discontinuities contribute to stability; and (6) the neglect of the crack propagation in rock bridges and the combined role of network connectivity and block interlocking. It should be emphasised that the resistance behaviour of rock bridges should not be factored in prematurely during the measurement process. Instead, rock-bridge measurement can be guided by the geometric descriptors described in Section 5.1, alongside the recording of stress conditions to facilitate subsequent mechanical analysis.

Regretfully, the mechanical analysis of rock bridges remains a significant challenge. Elmo (2023) clarified two fundamental restrictions to our understanding of rock bridges: (1) A rock bridge exists only at the moment it fails, and therefore we can neither see nor measure a rock bridge until it has failed; (2) Because of their conditional existence, we can only describe the contribution of rock bridges to rock mass strength as a potential. Note that this represents the mechanical measurement of rock bridges, rather than a geometric one. The rock bridges they refer to as being indescribable prior to failure in fact denote those whose resistance roles in failure processes cannot be determined. This introduces an open question in rock-bridge mechanics: which kinds of rock bridges, under what stress conditions, and to what extent they play resistant roles. Probabilistic and statistical methods may provide a preliminary solution, such as the stochastic modelling by Einstein et al. (1983) and the Bologna interpretation by Elmo (2023). Nevertheless, there is currently no definitive solution to this open question, as both rock-bridge configurations and mechanical behaviours are excessively complex; consequently, the measurement of mechanically significant rock bridges remains challenging.

## 6. Analytical method

As exemplified by the cross section in Fig. 1a, Terzaghi (1962) defined the area of the discontinuities after removing rock bridges as effective joint area, and defined the cohesion contributed by both discontinuities and rock bridges as effective cohesion $c_{\mathrm{eq}}$, expressed by:

$$c_{\mathrm{eq}} = c\frac{A_R}{A_D} \tag{11}$$

where $c$ is the cohesion of intact rocks, $A_R$ is the total area of rock bridges, and $A_D$ is the area of the cross section. This theoretical grounding catalysed further Lajtai's and Jennings's theories.

### 6.1. Lajtai's criterion

As rock bridges are frequently subjected to shear stresses in rock slopes and fault systems, Lajtai (1969c) proposed the strength criterion of rock bridges in direct shear, namely, Lajtai's criterion. He divided the failure mechanisms of intact rocks in direct shear into tension, shear, and failure at ultimate strength. Under a much lower confining stress, the minimum principal stress $\sigma_3$ is always in tension; therefore, failure in tension is the primary mechanism, and the failure envelope is expressed by the direct shear parabola:

$$\tau = \sqrt{T(T+\sigma_{\mathrm{n}})} \tag{12}$$

where $\sigma_{\mathrm{n}}$ and $\tau$ are the normal and shear stress components acting on the direct-shear plane, and $T$ is the tensile strength of intact rocks. As the confining stress increases, the failure envelope is defined by the Coulomb–Navier failure criterion:

$$\tau = \frac{1}{2}\sqrt{\frac{(2c+\sigma_{\mathrm{n}}\tan\phi)^2}{1+\tan^2\phi} - \sigma_{\mathrm{n}}^2} \tag{13}$$

where $\phi$ is the internal friction angle of intact rocks. As the confining stress increases sufficiently, the maximum direct-shear resistance is no longer governed by tensile or shear failure, but by frictional resistance of the crushed material. The ultimate-strength branch is therefore:

$$\tau = \sigma_{\mathrm{n}}\tan\phi_u \tag{14}$$

where $\phi_u$ is the angle of friction at ultimate strength.

Equations (12)–(14) are Lajtai's criterion for intact rocks in direct shear. Considering rock bridges in direct shear, stresses are solely transferred through rock bridges, and the stress components acting on rock bridges are magnified as:

$$\begin{cases} \sigma_{\mathrm{nr}} = \dfrac{\sigma_{\mathrm{n}}}{1-K} \\ \tau_r = \dfrac{\tau}{1-K} \end{cases} \tag{15}$$

where $\sigma_{\mathrm{n\,r}}$ and $\tau_{\mathrm{r}}$ are the normal and shear stress components acting on the rock bridge, and $K$ is the discontinuity persistence, which can be deduced by Eqs. (1)–(3). Substituting Eq. (15) to Eqs. (12)–(14), Lajtai's criterion for rock bridges in direct shear is expressed by:

$$\begin{cases} \tau = (1-K)\sqrt{T(T+\dfrac{\sigma_{\mathrm{n}}}{1-K})}\,\text{(Failure in tension)} \\ \tau = \dfrac{1}{2}\sqrt{\dfrac{[2(1-K)c+\sigma_{\mathrm{n}}\tan\phi]^2}{1+\tan^2\phi}-\sigma_{\mathrm{n}}^2}\,\text{(Failure in shear)} \\ \tau = \sigma_{\mathrm{n}}\tan\phi_u\,\text{(Failure at ultimate strength)} \end{cases} \tag{16}$$

For a closed discontinuity, normal stress is assumed to act across the entire plane, while the direct-shear resistance may be shared by rock bridges and frictional contact along the closed discontinuity segments. However, discontinuity friction may not be fully mobilised at the instant of rock-bridge failure. To account for this, Lajtai (1969a) introduced a mobilisation factor, here denoted by $m$. In this way, Eq. (16) becomes:

$$\begin{cases} \tau = mK\sigma_{\mathrm{n}}\tan\phi_{\mathrm{j}} + (1-K)\sqrt{T(T+\sigma_{\mathrm{n}})}\,\text{(Failure in tension)} \\ \tau = mK\sigma_{\mathrm{n}}\tan\phi_{\mathrm{j}} + \dfrac{1-K}{2}\sqrt{\dfrac{(2c+\sigma_{\mathrm{n}}\tan\phi)^2}{1+\tan^2\phi}-\sigma_{\mathrm{n}}^2}\,\text{(Failure in shear)} \\ \tau = \sigma_{\mathrm{n}}\tan\phi_u\,\text{(Failure at ultimate strength)} \end{cases} \tag{17}$$

where $\phi_{\mathrm{j}}$ is the internal friction angle of discontinuities.

### 6.2. Development of Lajtai's criterion

#### 6.2.1. Rock-bridge resistance model

Lajtai's criterion is proposed for collinear discontinuities; nevertheless, natural discontinuities may deviate from collinearity, such as those in rock slopes that may cause stepped failure surfaces. To this end, Einstein et al. (1983) modified Lajtai's criterion by introducing the rock-bridge resistance model. Given a shear direction along discontinuities $\beta$ and bridging angle $\alpha$, the rock-bridge resistance $F_{\mathrm{P}}$ can be expressed by:

$$F_P = \begin{cases} C\sqrt{T(T+\sigma_{\mathrm{n}})}, \alpha \le \theta_{\mathrm{t}} + \beta \\ Th, \alpha < \theta_{\mathrm{t}} + \beta \end{cases} \tag{18}$$

where $C$ is the continuity, $h$ is the discontinuity spacing, $\alpha$ is the bridging angle defined as $\sin^{-1}(h/L)$, and $\theta_{\mathrm{t}}$ is the angle of tensile cracks. Although this formula constrains the failure mechanism of rock bridges to tension and significantly simplifies the rock-bridge configuration, it highlights a critical shift: compared to shear strength, resistance is more applicable to engineering practice.

#### 6.2.2. Modification for the tensile failure criterion

In Lajtai's criterion, it is assumed that failure in tension occurs at discontinuity tips and rock bridges simultaneously. Nevertheless, observations show that failure occurs at discontinuity tips first. To fill in this gap, Liu and Xia (2010b) modified the strength criterion for tensile failure in Lajtai's criterion. As shown in Fig. 14a,b, assuming that the stress components of the rock-bridge element follow these distributions: $\sigma_y$ is uniformly distributed, while $\sigma_x$ and $\tau_{xy}$ are linearly distributed. The variation patterns of these elemental stress components are as follows:

$$\begin{cases} \sigma_x = \dfrac{\sigma_{\mathrm{m}} x}{L} \\ \sigma_y = \sigma_{\mathrm{n}} \\ \tau_{xy} = \dfrac{\sigma_{\mathrm{m}} y}{L} \end{cases} \tag{19}$$

where $\sigma_{\mathrm{m}}$ and $\sigma_{\mathrm{n}}$ are the stress components applied to the rock bridge, and $L$ is the length of the rock bridge. Given a plane that forms an angle to the direct-shear plane, the principal stresses can be expressed by:

$$\begin{cases} \sigma_1 = \dfrac{\sigma_{\mathrm{n}} + \sigma_x}{2} + \dfrac{1}{2}\sqrt{(\sigma_{\mathrm{n}} - \sigma_x)^2 + 4\tau_{xy}^2} \\ \sigma_3 = \dfrac{\sigma_{\mathrm{n}} + \sigma_x}{2} - \dfrac{1}{2}\sqrt{(\sigma_{\mathrm{n}} - \sigma_x)^2 + 4\tau_{xy}^2} \end{cases} \tag{20}$$

Assuming that the tensile failure of the rock bridge obeys the maximum tensile stress criterion, initial tensile failure occurs when $\sigma_3$ reaches $T$. In this way, the shear strength of the rock bridge can be expressed by:

$$\tau = \sqrt{(T - \sigma_x)(T - \sigma_{\mathrm{n}})} \tag{21}$$

Accordingly, at the right end of the rock bridge, $x = L$, and $\sigma_x = 0$. The stress upon tensile failure is:

$$\tau = \sqrt{T(T - \sigma_{\mathrm{n}})} \tag{22}$$

At the left end of the rock bridge, $x = 0$, and $\sigma_x = \sigma_{\mathrm{m}}$. The stress upon tensile failure is:

$$\tau = \sqrt{(T - \sigma_{\mathrm{m}})(T - \sigma_{\mathrm{n}})} \tag{23}$$

Therefore, the strength criterion for tensile failure in Lajtai's criterion is a special case where $\sigma_x = 0$. Equations (19)–(21) indicate that the shear stress levels follow an increasing gradient from the right discontinuity tip to the interior of the rock bridge, reaching a maximum at the left discontinuity tip.

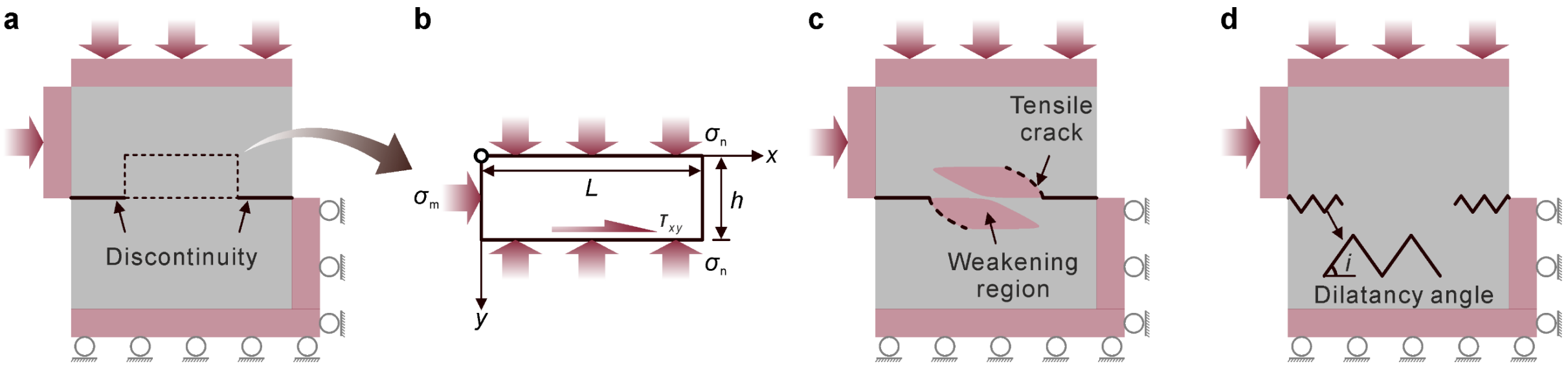

**Fig. 14. Modification for Lajtai's criterion. (a)** Coplanar rock bridge in direct shear. **(b)** Stress distribution at bridging regions. **(c)** Tensile cracks and weakening regions. **(d)** Rock bridge with regular serrations.

#### 6.2.3. Coupling with Barton's model

As Lajtai's criterion neglects the effects of discontinuity roughness, Wang et al. (2025b) coupled Lajtai's criterion with the strength criterion for discontinuities proposed by Barton and Choubey (1977) that is expressed as:

$$\tau = \sigma_{\mathrm{n}} \tan[\phi_{\mathrm{r}} + JRC \log(\frac{JCS}{\sigma_{\mathrm{n}}})] \tag{24}$$

where $\phi_{\mathrm{r}}$ is the residual friction angle, *JRC* is the joint roughness coefficient, and *JCS* is the discontinuity wall compressive strength. In this way, Eq. (16) becomes:

$$\begin{cases} \tau = Kc_{\mathrm{n}}\sigma_{\mathrm{n}} \tan[\phi_{\mathrm{r}} + \frac{1}{M'} JRC \log(\frac{JCS}{Kc_{\mathrm{n}}\sigma_{\mathrm{n}}})] + (1-K)\sqrt{T[T+(1-Kc_{\mathrm{n}})\sigma_{\mathrm{n}}]} \text{(Failure in tension)} \\ \tau = Kc_{\mathrm{n}}\sigma_{\mathrm{n}} \tan[\phi_{\mathrm{r}} + \frac{1}{M'} JRC \log(\frac{JCS}{Kc_{\mathrm{n}}\sigma_{\mathrm{n}}})] + \frac{1-K}{2}\sqrt{\frac{[2c+(1-Kc_{\mathrm{n}})\sigma_{\mathrm{n}}\tan\phi]^2}{1+\tan^2\phi} - (1-Kc_{\mathrm{n}})^2\sigma_{\mathrm{n}}^2} \text{(Failure in shear)} \end{cases} \tag{25}$$

where $c_{\mathrm{n}}$ is the transmission coefficient of the discontinuity that can be derived from the nonlinear regression by:

$$c_{\mathrm{n}} = 0.387\ln(K) + \frac{0.024\sigma_{\mathrm{c}}}{\sigma_{\mathrm{n}}} + 0.017JRC - 0.105 \pm 0.116 \tag{26}$$

where $\sigma_{\mathrm{c}}$ is the unconfined compressive strength, and *M'* is the damage degree of the joint that is given by Barton and Choubey (1977):

$$M' = \frac{JRC}{12\log(\frac{JCS}{Kc_{\mathrm{n}}\sigma_{\mathrm{n}}})} + 0.7 \tag{27}$$

### 6.3. Jennings's criterion

Equation (15) in Lajtai's criterion adopts the concept of a weighted average to calculate the stress components acting on a rock bridge. Following this concept, Jennings (1970) proposed a weighted average strength criterion for rock masses within the framework of the Mohr–Coulomb failure criterion, namely, Jennings's criterion:

$$\tau = K\left(c_{\mathrm{j}} + \sigma_{\mathrm{n}} \tan\phi_{\mathrm{j}}\right) + (1-K)\left(c + \sigma_{\mathrm{n}} \tan\phi\right) \tag{28}$$

where $c_{\mathrm{j}}$ is the cohesion of discontinuities. In this manner, the unconfined compressive strength $\sigma_{\mathrm{c}}$ can be calculated as:

$$\sigma_{\mathrm{c}} = \frac{2c_{\mathrm{eq}}\cos\phi_{\mathrm{eq}}}{1-\sin\phi_{\mathrm{eq}}} \tag{29}$$

where $c_{\mathrm{eq}}$ is the equivalent cohesion, and $\phi_{\mathrm{eq}}$ is the equivalent friction angle. They are defined by:

$$\begin{cases} c_{\text{eq}} = (1-K)c + Kc_{\text{j}} \\ \phi_{\text{eq}} = \tan^{-1}\left[(1-K)\tan\phi + K\tan\phi_{\text{j}}\right] \end{cases} \tag{30}$$

Due to the simplicity of Jennings's criterion, numerous case studies have employed it in stability analysis of rock slopes, such as the right abutment of the Upper Gotvand Dam in Iran (Broojerdi et al., 2018), 2018 slope failure in the Serra do Sapo open pit mine in Brazil (Figueiredo et al., 2021), and Drus potential collapse in France (Matasci et al., 2015).

### 6.4. Development of Jennings's criterion

#### 6.4.1. Rock-bridge weakening model

In direct shear tests on rock bridges, Liu and Xia (2010a) observed that rock-bridge failure is both localised and progressive. Localisation refers to the non-uniform occurrence of failure, which concentrates in specific bridging regions. Progressivity denotes the weakening of mechanical properties of rock bridges prior to macroscopic rupture. As shear stress increases, microscopic cracks develop, and macroscopic tensile cracks initiate and propagate into the interior of the rock bridge, ultimately leading to instability (Fig. 14c).

Specifically, Xia et al. (2010) noticed that tensile stress concentration zones form in the lower-front region of the left joint tip and the upper-front region of the right joint tip (Fig. 14c), triggering tensile failure and the formation of wing cracks. These wing cracks propagate rapidly toward the direction of $\sigma_1$ before quickly stabilising. As shear stress further increases, the extent of microcracking within the stress concentration zones expands, leading to specimen dilation and the resultant dilatancy effect. Under this effect, the influence of normal stress weakens, and the shear strength at the rock bridge decreases, thereby intensifying the shear stress in the rock bridge. Shear cracks then form within the rock bridge along the shearing direction. Only upon reaching the peak shear strength do the internal shear cracks in the rock bridge interconnect with the joints on both sides.

The weakening degree $\delta$ is defined as the ratio of failed rock-bridge elements $N_{\text{t}}$ to the total number of elements $N$. Given that the number of failed elements is correlated with $\sigma_3$, the $\delta$ expressed as a function of the $\sigma_3$ of rock-bridge elements is given as (Stimpson, 1978):

$$\delta = \frac{N_{\text{t}}(\sigma_3)}{N} = 1 - \exp\left[-\left(\frac{\sigma_3}{\sigma_0}\right)^{m_0}\right] \tag{31}$$

Based on the relationship between the stress state of rock-bridge elements and shear displacement, Xia et al. (2010) proposed a $\delta$ formula using shear displacement as the primary variable:

$$\delta = 1 - \exp\left[ -\left( \frac{u - u_0}{u_\mathrm{f}} \right)^{m_0} \right] \tag{32}$$

where $u$ is the shear displacement, $u_0$ is the sliding displacement of discontinuities, $u_\mathrm{f}$ is the shear displacement upon failure, $m_0$ is a parameter to quantify the effect of rock material on $\delta$. The shear strength of the rock mass is composed of three parts: (1) the shear strength of the unweakened segments of the rock bridge; (2) the shear strength of the weakened segments; and (3) the shear strength of the discontinuity. In this way, Jennings' criterion becomes (Tang et al., 2012):

$$\tau = K\left(c_\mathrm{j} + \sigma_\mathrm{n} \tan\phi_\mathrm{j}\right) + (1-\delta)(1-K)\left(c + \sigma_\mathrm{n} \tan\phi\right) + \delta(1-K)(1-\frac{\delta}{m_0})\left(c + \sigma_\mathrm{n} \tan\phi\right) \tag{33}$$

Despite incorporating rock-bridge weakening, this model is restricted to direct shear tests. As noted by Tang et al. (2012), several parameters rely on empirical data, and the averaging approach for calculating weakening is overly simplified, thereby imposing significant limitations on its application.

#### 6.4.2. Dilatancy angle model

The roughness of discontinuities significantly influences their shear strength. To incorporate this effect, Xia et al. (2010) derived the modified Jennings's criterion, which accounts for the dilatancy angle $i$ (Fig. 14d). In the shear strength formula for persistent discontinuities, the friction angle comprises two components: the basic friction angle and the asperity overriding angle caused by discontinuity roughness. Correspondingly, during the shearing of a rock bridge, the friction angle of the portion of discontinuities $\phi_\mathrm{jD}$ should also consist of the basic friction angle $\phi_\mathrm{j}$ and the overriding angle of regular serrations $i$:

$$\phi_\mathrm{jD} = \phi_\mathrm{j} + i \tag{34}$$

Since the influence of discontinuity undulation on shear strength is dependent on the normal stress on the shear plane, Eq. (30) can be modified as follows:

$$\begin{cases} c_\mathrm{eq} = (1-K)c + Kc_\mathrm{j} \\ \tan\phi_\mathrm{eq} = (1-K)\tan\phi + K\tan(\phi_\mathrm{j} + i \cdot f) \end{cases} \tag{35}$$

where $f$ is a variable related to the discontinuity properties. Substituting Eq. (35) into Eq. (28), Jennings's criterion is modified as:

$$\tau = K[c_\mathrm{j} + \sigma_\mathrm{n} \tan(\phi_\mathrm{j} + i \cdot f)] + (1-K)\left(c + \sigma_\mathrm{n} \tan\phi\right) \tag{36}$$

Equation (36) can also incorporate the rock-bridge weakening model in Eq. (33), and Jennings's criterion becomes:

$$\tau = K[c_\mathrm{j} + \sigma_\mathrm{n} \tan(\phi_\mathrm{j} + i \cdot f)] + (1-\delta)(1-K)\left(c + \sigma_\mathrm{n} \tan\phi\right) + \delta(1-K)(1-\frac{\delta}{m_0})\left(c + \sigma_\mathrm{n} \tan\phi\right) \tag{37}$$

Several studies have applied similar principles to modify Jennings's criterion (Hu et al., 2020; Zheng et al.,

2020; Yi and Xu, 2022), though they are not detailed here.

### 6.4.3. Coupling with Jaeger's model

In Jennings's criterion, given identical discontinuity arrangements subjected to different stress directions, the mechanical responses of rock masses are the same, which conflicts with the fact of stress-orientation dependence. Rock mass strength presents anisotropy. To quantify this, Jaeger et al. (1969) proposed the single plane of weakness theory. For an inclined discontinuity oriented at an angle $\beta$ to $\sigma_3$ under uniform compressive far-field stress for plane strain problems, the normal and shear stress acting along the discontinuity can be expressed as:

$$\begin{cases} \sigma_{\mathrm{nj}} = \dfrac{\sigma_1 + \sigma_3}{2} + \dfrac{(\sigma_1 - \sigma_3)\cos(2\beta)}{2} \\ \tau_{\mathrm{j}} = \dfrac{\sigma_1 - \sigma_3}{2}\sin(2\beta) \end{cases} \tag{38}$$

Complying with the Coulomb friction criterion:

$$\tau_{\mathrm{j}} = c_{\mathrm{j}} + \sigma_{\mathrm{nj}} \tan\phi_{\mathrm{j}} \tag{39}$$

The condition for the shear failure along the discontinuity can be formulated as:

$$\sigma_1 = \sigma_3 + \frac{2(c_{\mathrm{j}} + \sigma_3 \tan\phi_{\mathrm{j}})}{(1 - \cot\beta \tan\phi_{\mathrm{j}})\sin(2\beta)} \tag{40}$$

Equation (40) has two intersections with the Mohr's circle of intact rocks; the corresponding $\beta$ solutions are denoted as $\beta_1$ and $\beta_2$. When $\beta$ falls in the range [$\beta_1$, $\beta_2$], rupture occurs along the discontinuity; otherwise, it follows the classical failure plane with $\theta$ = 45° + $\phi$/2. Rafiei Renani et al. (2019) have attempted to couple Eq. (40) with Jennings's criterion, where rock bridges are treated as equivalent discontinuities with enhanced cohesion and friction; thus, $c_{\mathrm{j}}$ and $\phi_{\mathrm{j}}$ in Eq. (40) are replaced by $c_{\mathrm{eq}}$ and $\phi_{\mathrm{eq}}$ in Eq. (30), respectively. Similarly, failure occurs along rock bridges and discontinuities when $\beta$ lies within the range [$\beta_{1\mathrm{eq}}$, $\beta_{2\mathrm{eq}}$]. According to the sine rule on the Mohr's circle, $\theta_{1\mathrm{eq}}$ and $\theta_{2\mathrm{eq}}$ can be expressed as:

$$\begin{cases} \beta_{1\mathrm{eq}} = \dfrac{\phi_{\mathrm{eq}}}{2} + \dfrac{1}{2}\sin^{-1}\left[\dfrac{(\sigma_1 + \sigma_3 + 2c_{\mathrm{eq}}\cot\phi_{\mathrm{eq}})\sin\phi_{\mathrm{eq}}}{\sigma_1 - \sigma_3}\right] \\ \beta_{2\mathrm{eq}} = \dfrac{\pi}{2} + \dfrac{\phi_{\mathrm{eq}}}{2} - \dfrac{1}{2}\sin^{-1}\left[\dfrac{(\sigma_1 + \sigma_3 + 2c_{\mathrm{eq}}\cot\phi_{\mathrm{eq}})\sin\phi_{\mathrm{eq}}}{\sigma_1 - \sigma_3}\right] \end{cases} \tag{41}$$

## 6.5. Linear elastic fracture mechanics (LEFM) models

### 6.5.1. Mode II fracture by Hung and Lee (1990)

Given a plane composed of mesoscopic discontinuities and a rock bridge, Hung and Lee (1990) derived its shear

strength:

$$\tau = \tau_{\mathrm{s}} + \frac{K_{\mathrm{IIC}}}{\sqrt{L_c}F(p)} \tag{42}$$

where $\tau_{\mathrm{s}}$ is the shear strength of the discontinuity, $F(p)$ is the dimensionless function that maps model geometry to external loading types, calibrated via curve-fitting, $L_c$ is the length of impersistent discontinuity, and $K_{\mathrm{IIC}}$ is the Mode II fracture toughness of the rock determined by:

$$K_{\mathrm{IIC}} = \sqrt{\frac{G_{\mathrm{IIC}}E}{1-\nu^2}} \tag{43}$$

where $G_{\mathrm{IIC}}$ is the Mode II fracture energy, $E$ is the Young's modulus, and $v$ is the Poisson's ratio.

### 6.5.2. Mode II fracture by Xu and Zhu (1995)

To account for stress concentration at discontinuity tips, Xu and Zhu (1995) proposed the rock-bridge shear strength formula based on fracture mechanics:

$$\tau = \frac{K_{\mathrm{IIC}}}{\sqrt{2b\tan(\pi a/2b)}} + \sigma_{\mathrm{n}}\tan\phi_{\mathrm{j}} + c_{\mathrm{j}} \tag{44}$$

where $2a$ represents the discontinuity length, $2b$ denotes the distance between the centres of two adjacent discontinuities, and $K_{\mathrm{IIC}}$ is the Mode II fracture toughness of the rock.

Ren et al. (2003) noted that Eq. (44) corresponded to the tensile crack initiation. According to their experiments, following tensile crack initiation and propagation, shear cracks initiated, propagated, and connected the tensile cracks, resulting in rock-bridge failure. Accordingly, Eq. (44) becomes:

$$\tau(\beta_{\mathrm{s}}) = \left[\frac{(b-ac_{\mathrm{n}})}{b}\tan(\beta_{\mathrm{s}}+\phi_0') + \frac{a}{b}c_{\mathrm{n}}\tan\phi_{\mathrm{j}}\right]\sigma_{\mathrm{n}} + \frac{(b-a)c_0'\cos\phi_0'\sin\alpha'}{b\cos(\beta_{\mathrm{s}}+\phi_0')\sin(\beta_{\mathrm{s}}+\alpha')} + \frac{a}{b}c_{\mathrm{j}} \tag{45}$$

where $\beta_{\mathrm{s}}$ is the inclination of the shear crack relative to the shear direction, $c_{\mathrm{n}}$ is the transmission coefficient of the discontinuity, $\phi_0'$ is the weakened internal friction angle of intact rocks, $c_0'$ is the weakened cohesion of intact rocks, $\alpha'$ is the angle between the discontinuity and $\sigma_1$.

It is evident that Eq. (45) contains several parameters that are difficult to obtain, such as $\phi_0'$ and $c_0'$. In addition, it can only be applied to specific confining stress conditions under direct shear, as the failure mechanism may be different under alternative confining stress states.

### 6.5.3. Mode I fracture by Kemeny and Cook (1986)

For non-interacting circular cracks with radius, Kemeny and Cook (1986) formulated the tensile strength $T$ as a function of the cracked cross-sectional area percent, specifically for cracks oriented perpendicular to the loading

axis:

$$T = \frac{\pi}{2} K_{\mathrm{IC}} \sqrt{\frac{N^{1/3}}{\sqrt{\pi K}}} \text{ for } K \ll 1 \tag{46}$$

where $N$ is the number of cracks in a unit volume, and $K_{\mathrm{IC}}$ is the Mode I fracture toughness of the rock. It can also be expressed by the rock-bridge percent:

$$T = \frac{\pi}{2} K_{\mathrm{IC}} \sqrt{\frac{(1-K)^{3/2} N^{1/3}}{\sqrt{\pi}}} \text{ for } (1-K) \ll 1 \tag{47}$$

### 6.5.4. Mode I fracture by Bagheripour (1996)

Bagheripour (1996) characterised rock-bridge fracturing using the maximum tangential stress (MTS) criteria (Erdogan and Sih, 1963):

$$\sigma_1 = \sigma_3 + \frac{\sqrt{3/a} K_{\mathrm{IC}} / f_{cr} + 2\sigma_3 \tan\phi_{\mathrm{j}}}{(1-\cot\beta \tan\phi_{\mathrm{j}})\sin(2\beta)} \tag{48}$$

where $f_{\mathrm{cr}}$ is a correction factor introduced to account for discontinuity length. It significantly deviates from the experimental results.

## 6.6. Elastic-plastic fracture mechanics (EPFM) models

Zhou et al. (2026) characterised rock-bridge fracturing using the maximum plastic zone fracture criterion. They derived the plastic zone radii at flaw tips to determine the initiation and propagation of wing cracks, and the contact of plastic zones serves as a proxy for rock-bridge failure. The plastic zone radius $r$ at flaw tips can be determined as:

$$r = \frac{1}{2\pi T^2} \left[ \kappa K_{\mathrm{I}} \cos\frac{\theta_{\mathrm{a}}}{2} - \kappa K_{\mathrm{II}} \sin\frac{\theta_{\mathrm{a}}}{2} + \frac{1}{2} \sqrt{K_{\mathrm{I}}^2 \sin^2\theta_{\mathrm{a}} + 2K_{\mathrm{I}} K_{\mathrm{II}} \sin 2\theta_{\mathrm{a}} + K_{\mathrm{II}}^2 (4 - 3\sin^2\theta_{\mathrm{a}})} \right]^2 \tag{49}$$

where $\kappa = 1-2\nu T/\sigma_{\mathrm{c}}$ for plane strain problem, $\kappa = 0$ for plane stress problem, $K_{\mathrm{I}}$ is the stress intensity factor (SIF) of type I, $K_{\mathrm{II}}$ is the SIF of type II, and $\theta_{\mathrm{a}}$ is the arbitrary angle direction. Further details of the solution to $K_{\mathrm{I}}$, $K_{\mathrm{II}}$, and $\theta_{\mathrm{a}}$ can be found in Zhou et al. (2026).

## 6.7. Theoretical impasse

Lajtai's criterion and Jennings's criterion, alongside fracture mechanics frameworks, remain the dominant paradigms in the rock-bridge domain. While refined over decades and integrated into various models, their fundamental deficiencies persist.

Both Lajtai's criterion and Jennings's criterion are based on the concept of discontinuity persistence *K*, which has been acknowledged to possess limitations (ISRM, 1978), such as its restricted applicability to coplanarity (Einstein et al., 1983), neglect of progressive failure (Elmo et al., 2018), and omission of scale effects (Bu et al., 2022). In addition, Lajtai's criterion, developed from direct shear tests, has not been generalised to alternative stress states. Subsequent refinements remain constrained by the concept of *K* and direct shear tests. Furthermore, the discrete consideration of the three failure modes precludes its application to the mixed-mode failure prevalent in nature and experiments, thereby limiting its practical adoption. In contrast, Jennings's criterion focuses exclusively on shear failure, failing to account for tensile and mixed-mode failure mechanisms common in nature and experiments. Furthermore, it suffers from a heavy reliance on the concept of *K* and overlooks the stress-orientation dependence of rock-bridge behaviour. Subsequent modifications, while incorporating progressive fracturing, have failed to overcome these fundamental limitations.

Given the stress concentrations at rock bridges due to the significant contrast in their mechanical properties than those of discontinuities, many researchers have sought solutions within fracture mechanics. However, whether employing Mode I/II criteria under LEFM or under EPFM frameworks, several fundamental deficiencies persist: (1) Fracture mechanics describes crack initiation and propagation; however, it has difficulty describing crack coalescence, which is the primary process of rock-bridge failure, as crack coalescence is a highly nonlinear complex mechanical process wherein cracks interact and microcracks connect in intractable ways (Segall and Pollard, 1980; Kemeny and Cook, 1986; Wong and Einstein, 2009a; Fossen and Rotevatn, 2016; Einstein, 2021). In simpler terms, fracture mechanics fails to interpret how a rock bridge resists rupture as a whole. (2) While they may be accurate for individual cracks like wing cracks, both LEFM and EPFM criteria should be changed accordingly when the number of discontinuities and their arrangements are increased or changed, making them difficult to describe complex, multi-modal failure processes. (3) The excessive reliance on burdensome parameters and assumptions. These fundamental deficiencies would introduce significant uncertainty when interpreting the mechanical behaviour of rock bridges.

## 7. Rock mechanics research

Figure 3e indicates that fundamental research on rock bridges accounts for a substantial proportion. Most of them are based on rock mechanics, and attempts to investigate the strength, failure mode, cracking process, and determinant, under different stress regimes, employing theoretical, experimental, numerical, and field-based approaches. Following the previous review of theoretical and field-based studies of rock bridges, this section focuses on experimental and numerical results, categorised by stress regimes.

### 7.1. Rock-bridge failure in unconfined/confined shear

Shear failure is one of the primary failure mechanisms of rock bridges. Given that shear stress states are prevalent in landslide sliding surfaces and seismic faulting, the mechanical behaviour of rock bridges under shear loading has been extensively investigated. Lajtai (1969c) pioneered direct shear tests on rock bridges. He conducted direct shear tests on gypsum specimens containing two collinear discontinuities under various confining stresses, identifying three distinct failure modes of rock bridges under direct shear: failure in tension, failure in shear, and failure at ultimate strength. Following this experimental setup, Savilahti et al. (1991) conducted direct shear tests on gypsum specimens containing collinear and 45°-inclined discontinuities under relatively low confining stresses, and concluded that rock bridges often fail in tension. Another representative early work is from Gehle and Kutter (2003), who conducted 131 direct shear tests on gypsum and limestone specimens containing en échelon discontinuity arrays. They found that rock-bridge failure in direct shear has multiple mechanisms and phases. It initiates by wing-crack propagation from pre-existing discontinuities into the rock bridges, then culminates in the generation of secondary fractures that interconnect the initial flaws. They also noted that the strength of impersistent discontinuities is governed by a composite of several distinct mechanical processes rather than a solitary mechanism. Many extant models for evaluating the shear strength of jointed rock masses typically rely on a single-mechanism framework or are purely empirical; consequently, they are inadequate for representing the comprehensive evolution of the shear process from incipient fracture to residual sliding.

Further rock mechanics studies have demonstrated that the complexity of the mechanical behaviour of rock bridges in direct shear lies in rock-bridge configurations and boundary conditions. For the rock-bridge configurations, various discontinuity arrangements have been pre-fabricated to study their possible effects, including discontinuity length (Wong et al., 2001a; Alzo'ubi et al., 2007; Ghazvinian et al., 2011; Asadizadeh et al., 2018; Yang and Kulatilake, 2019; Sun et al., 2020; Tang et al., 2020; Chen et al., 2021b; Fereshtenejad et al., 2021; Yang et al., 2022b; Qin et al., 2023; Zhang et al., 2024; Zhang et al., 2025; Le et al., 2026; Luo et al., 2026), discontinuity orientation (Zhang et al., 2006; Jiang et al., 2017; Cao et al., 2018; Zhang et al., 2020b; Yin et al., 2021; Al-E'Bayat et al., 2023), discontinuity spacing (Wong et al., 2001a; Ghazvinian et al., 2011; Sarfarazi et al., 2014; Asadizadeh et al., 2018; Zhang et al., 2024), discontinuity aperture (Meng et al., 2022), discontinuity overlap (Ghazvinian et al., 2011; Sarfarazi et al., 2014; Haeri, 2015; Rajput et al., 2022), discontinuity roughness (Asadizadeh et al., 2018; Fereshtenejad et al., 2021; Xue et al., 2024), and discontinuity intensity (Zhang et al., 2020a; Bu et al., 2022) (Fig. 15). For example, for discontinuity length, whether utilising natural rock specimens (e.g. granite (Yang and Kulatilake, 2019; Luo et al., 2026) and sandstone (Sun et al., 2020; Yang et al., 2022b; Qin et al., 2023; Zhang et al., 2025)), rock-like analogues (Wong et al., 2001a; Alzo'ubi et al., 2007; Ghazvinian et al., 2011; Asadizadeh et al., 2018; Tang et al., 2020; Fereshtenejad et al., 2021), or numerical simulations (Ghazvinian et al., 2011; Chen et al., 2021b; Zhang et al., 2024), an expected result is always derived: the maximum load sustained by the specimen containing rock bridges decreases as the discontinuity length increases. Rock-bridge fracturing may also be affected. For discontinuity orientation, it profoundly dictates rock mass strength and rock-bridge fracturing processes. While an increase in orientation can enhance the shear strength of the rock mass and induce a failure mode transition from tensile to shear, it may also yield diametrically opposite outcomes depending on the context (Zhang et al., 2006; Jiang et al., 2017; Cao et al., 2018; Zhang et al., 2020b; Yin et al., 2021; Al-E'Bayat et al., 2023). Consequently, discontinuity orientation remains one of the most intricate parameters governing rock-bridge behaviour, a complexity fundamentally rooted in the acute sensitivity of the rock-bridge configurations to discontinuity arrangements. Variations in one or more discontinuity orientations commensurately alter the rock-bridge configuration. Most studies have failed to isolate variables according to the geometric descriptors defined in Section 5.1. Consequently, manipulated discontinuity orientations often conflate multiple geometric descriptors of rock bridges, introducing significant uncertainty into the reported findings.

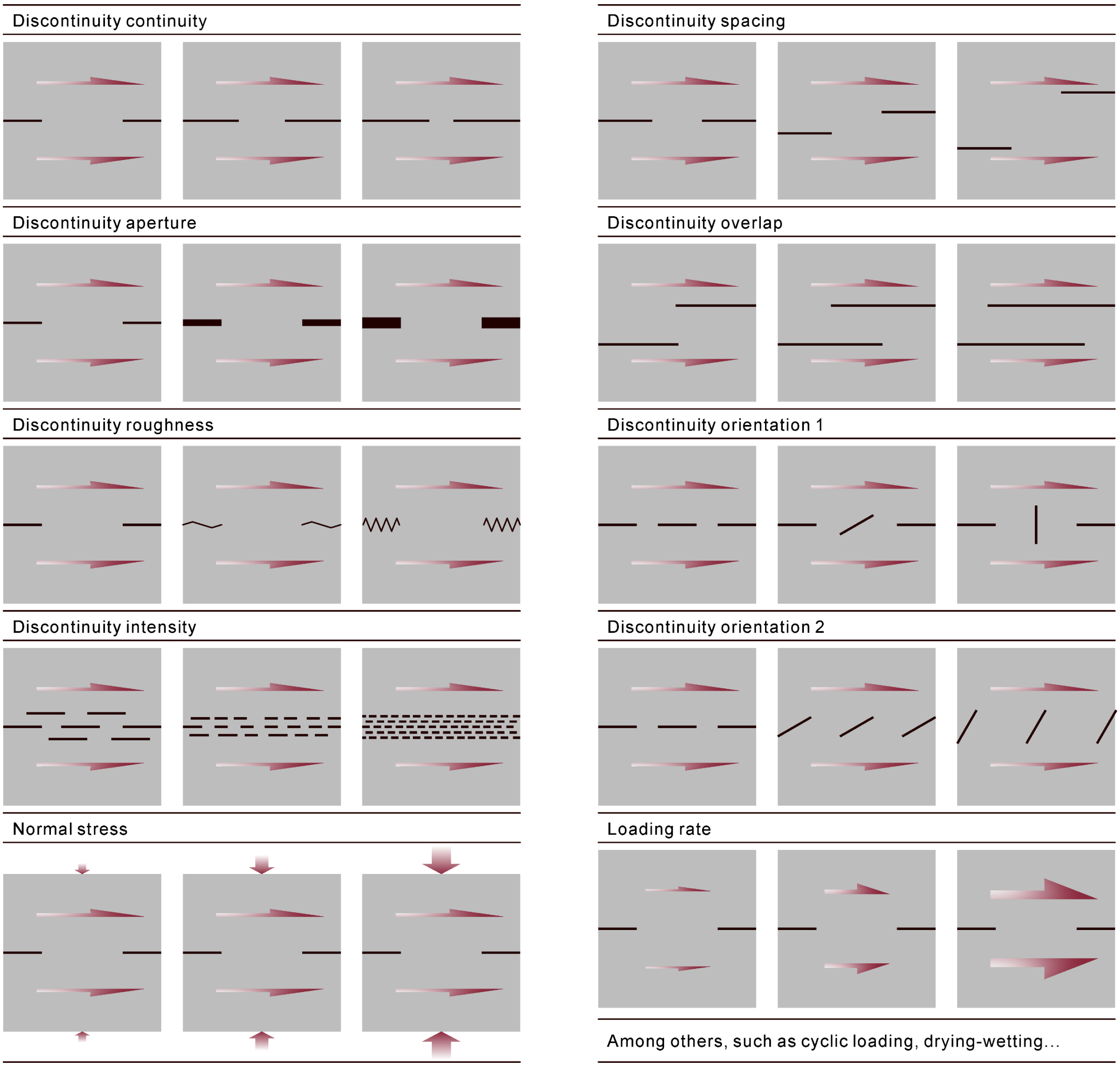

**Fig. 15. Various research scenarios in direct shear tests on rock bridges.**

The effects of boundary conditions include normal stress and loading rate. Normal stress is almost ubiquitously adopted as a primary parameter in direct shear testing of rock bridges (Lajtai, 1969c; Yang and Kulatilake, 2019; Sun et al., 2020; Chen et al., 2021b; Fereshtenejad et al., 2021; Rajput et al., 2022; Yang et al., 2022b; Qin et al., 2023; Miao et al., 2024). Under a low normal stress, rock-bridge failure is dominated by tensile fracturing; conversely, under a high normal stress, the failure mechanism transitions primarily to shear. Under low normal stress, incipient fractures propagate heterogeneously via paths of least resistance, whereas elevated stress constrains expansion to more uniform trajectories. For loading rate, experimental and numerical investigations have shown that rock-bridge failure is governed by curved tensile fractures at low shear rates, whereas elevated shear rates drive a transition towards linear shear fracturing (Luo et al., 2022). The shear strength also increases with increasing loading rate, consistent with that in unconfined/confined compressive tests. Other reported boundary conditions that may affect rock-bridge behaviour include hydraulic pressure (Yang et al., 2022b; Yang et al., 2025), cyclic loading (Wang et al., 2025a), etc.

These findings demonstrate that failure processes in rock bridges in direct shear involve synergistic factors that elude the descriptive capacities of existing models including Lajtai's criterion, Jennings's criterion, and fracture mechanics frameworks. Consequently, our understanding of rock-bridge behaviour remains restricted to phenomenological or statistical assessments, leaving the fundamental rock-bridge mechanics poorly defined. Mapping these insights to complex *in situ* fracture networks is even more challenging, given that a reliable framework of rock-bridge mechanics to constrain results even under controlled laboratory conditions has yet to be established.

### 7.2. Rock-bridge failure in unconfined/confined compression

Natural rock bridges are frequently subjected to compressive stress states. Unconfined/confined compressive tests in a lab or simulation can simulate such stress states. Therefore, there are numerous studies of rock-bridge failure in unconfined/confined compression. Brace and Bombolakis (1963) pioneered unconfined compressive tests on PMMA containing en échelon discontinuities. Tensile cracks initiated at discontinuity tips and propagated along the axial loading direction. However, rock-bridge failure was not presented. Horii and Nemat-Nasser (1985) conducted unconfined/confined compressive tests on Columbia Resin CR-39, characterised by either single-row or multi-row configurations of short flaws combined with longer flaws. Five coalescence patterns (rock-bridge failure modes) were observed. Reyes and Einstein (1990) conducted unconfined compressive tests on gypsum specimens containing two inclined offset joints. They demonstrated that fracture propagation is fundamentally governed by joint geometry, with initial wing crack development preceding the growth of secondary cracks that drive coalescence in non-overlapping configurations. Conversely, for overlapping flaws, coalescence occurred through tensile wing crack extension at a significantly higher load threshold than that observed for non-overlapping cases.

Given the complexity of crack coalescence in rock bridges, researchers began interpreting this process through a taxonomic lens, which has since become the dominant paradigm in fundamental research on rock bridges. Shen et al. (1995) conducted unconfined compressive tests on gypsum specimens containing two parallel and offset joints. They identified thirteen types of rock-bridge failure. With fixed ligament length and variant bridging angle, there was a general trend in the variation of rock-bridge failure modes: with a small positive bridging angle and negative bridging angle, the coalescence was mainly achieved by shear cracks linking up the inner discontinuity tips; with an intermediate bridging angle, the coalescence was mainly achieved by shear cracks and tensile cracks; with a large bridging angle, the coalescence was mainly achieved by tensile cracks. The coalescence featuring shear cracks always initiated at discontinuity tips and propagated towards the centre of rock bridges; while the coalescence featuring the combination of shear and tensile cracks always initiated at the centre of rock bridges and propagated towards discontinuity tips.

Bobet and Einstein (1998) further extended the work of Shen et al. (1995) by introducing biaxial compressive tests. They conducted both uniaxial and biaxial compressive tests on gypsum rock bridges containing two open or closed offset joints. Their observations revealed that tensile wing cracks, initiating at flaw tips, were the first fractures to emerge under loading. These were followed by secondary cracks, which typically developed in a coplanar orientation to the pre-existing discontinuities. Morphological evidence, including localised spalling and the presence of crushed gypsum powder on the fracture surfaces, led to the classification of these secondary features as shear cracks. Consequently, the study categorised all shear-driven fractures as secondary cracks and all tensile-driven features as wing cracks. Accordingly, they classified the observed rock-bridge failure into five modes (Fig. 16): (1) Secondary shear crack, initiated at discontinuity tips; and failure surface was rough and contained many small kink steps and crushed gypsum; (2) Secondary shear and tensile cracks, initiated at discontinuity tips; some failure surface were clean and smooth, while some were rough with crushed gypsum; (3) Secondary shear crack and wing crack, initiated at discontinuity tips; some failure surface were clean and smooth, while some were rough with crushed gypsum; (4) Wing crack, initiated at discontinuity tips; and failure surface was smooth and clean; (5) Secondary crack, initiated at discontinuity tips; and failure surface was very rough with a lot of crushed gypsum. In the same year, Wong and Chau (1998) observed three rock-bridge failure modes: (1) Shear; (2) Mixed shear-tension; and (3) Wing tension. The mutual combination of these three modes yields nine configurations that elucidate the observed failure morphologies (Fig. 16). Subsequently, the work of Bobet and Einstein (1998) was further extended by Sagong and Bobet (2002), Park and Bobet (2009), and Ko and Lee (2020), where the classification system was further improved.

High-speed cameras have enabled a more precise characterisation of rock-bridge failure. Martinez (1999) used a high-speed camera up to 250 FPS to record the details of rock-bridge failure. He replaced the gypsum material in the work of Bobet and Einstein (1998) with Vermont white marble and Barre granite. He conducted unconfined compressive tests on rock bridges containing two offset joints. Five coalescence modes were identified: (1) Linkage of two internal shear cracks; (2) Linkage of two internal shear cracks by a vertical tensile crack (not a wing crack); (3) Propagation of the internal shear crack from one of the flaws until it reaches the internal wing crack of the other flaw; (4) Propagation of an internal wing crack from one flaw until it reaches the other flaw. Observed only for granite, but not for marble; (5) Linkage of two internal wing cracks that propagate until they join each other halfway. Observed only for granite, but not for marble. Wong and Einstein (2009a; 2009b) further extended this work. They conducted unconfined compressive tests on gypsum and Carrara marble rock bridges containing two offset joints. Before classifying rock-bridge failure modes, they clarified that primary and secondary cracks are classified chronologically, whereas shear and tension are categorised by mechanism, suggesting that the latter provides a clearer understanding of the failure process of rock bridges. According to the video recorded by the high-speed camera during the experiment, they identified nine crack coalescence modes (Fig. 16): (1) No coalescence; (2) Indirect coalescence by two or multiple cracks (crack types vary); (3) Type 2 shear crack(s); (4) Type 1 shear crack(s); (5) One or more type 2 shear crack(s) and type 2 tensile crack segments between inner flaw tips; (6) Type 2 tensile crack(s). There may be occasional short shear segments present along the coalescence crack. (7) Type 1 tensile crack(s); (8) Flaw tips of the same side linked up by tensile crack(s) not displaying wing appearance (crack type not classified). There may be occasional short shear segments present along the coalescence crack; (9) Type 3 tensile crack(s) linking the right tip of the top flaw and left tip of the bottom flaw. There may be occasional short shear segments present along the coalescence crack. In this classification system, crack types are defined by Wong and Einstein (2009a; 2009b).

More refined observational and analysis techniques have been introduced into the monitoring of rock-bridge fracturing, such as CT and numerical simulation. For example, Yang et al. (2022a) conducted confined compressive tests on sandstones containing multiple offset joints. According to the CT results, they obtained the internal rock-bridge fracture characteristics. Accordingly, they identified five crack coalescence patterns: (1) One wing crack (or secondary crack) and one anti-wing crack connect with each other. (2) A shear crack links the two discontinuity tips directly. (3) An axial tensile crack coalesces directly between the inner discontinuity tips. (4) An axial tensile crack links the outer tip of one discontinuity and the inner tip of the other discontinuity. At the same time, one tensile crack initiates from the inner tip of the latter discontinuity and propagates towards the bottom of the specimen. (5) Two axial tensile cracks link the two discontinuity tips (Fig. 16). They also revealed the effect of confining pressure on fracturing processes. Zhang and Wong (2013) extended the work of Wong and Einstein (2009a) by conducting DEM simulations of unconfined compressive tests on gypsum rock bridges containing two offset joints. Two additional types of coalescence patterns were identified: (1) A number of steeply inclined to vertical macro-tensile cracks; and (2) Type 1 tensile crack(s) and mixed tensile-shear crack initiated from two flaws, respectively. Zhou et al. (2015) employed GPD simulations. Results revealed four coalescence mechanisms: (1) Tension; (2) Compression; (3) Shear; and (4) Mixed tension/shear. Shear failure required a higher load compared to tension and mixed tension/shear.

Subsequent studies have some similarities and some supplementary with previous classifications, such as that conducted by Zhou et al. (2024). They noted that differences in loading conditions and discontinuity geometry may lead to variations in the coalescence types summarised.

The abovementioned findings indicate that rock-bridge failure under compression, featuring crack coalescence, is a highly complex, multifactorial process. Similar to the direct shear conditions detailed in Section 7.1, these influence factors can be divided into rock-bridge configurations and boundary conditions. Rock-bridge configurations are acutely sensitive to the discontinuity arrangements, such as geometry (Yu et al., 2020; Wang et al., 2023), aperture (Wang et al., 2020b), and friction (Wong and Chau, 1998; Wong et al., 2001b), where the convoluted distribution of these features (Figs. 15 and 16) directly dictates the intricacy of the coalescence process. Boundary conditions encompasses broader variables, such as loading rate (Wu et al., 2019), freeze-thaw cycle (Zhou et al., 2022a; Deng et al., 2024), confining pressure (Wu et al., 2019; Yang et al., 2022a), drying-wetting: (Zheng et al., 2024), and cyclic load (Xu et al., 2021; Zhou et al., 2022a; Deng et al., 2026), which further modulate coalescence processes. Consequently, current research is largely restricted to the taxonomic classification of observed crack coalescence patterns; however, these systems require continual refinement, as the sheer diversity of coalescence pathways necessitates frequent recalibration of established frameworks. The infinite permutations of discontinuity arrangements and boundary conditions merely exacerbate taxonomic complexity, offering little insight into the fundamental mechanical essence of rock bridges.

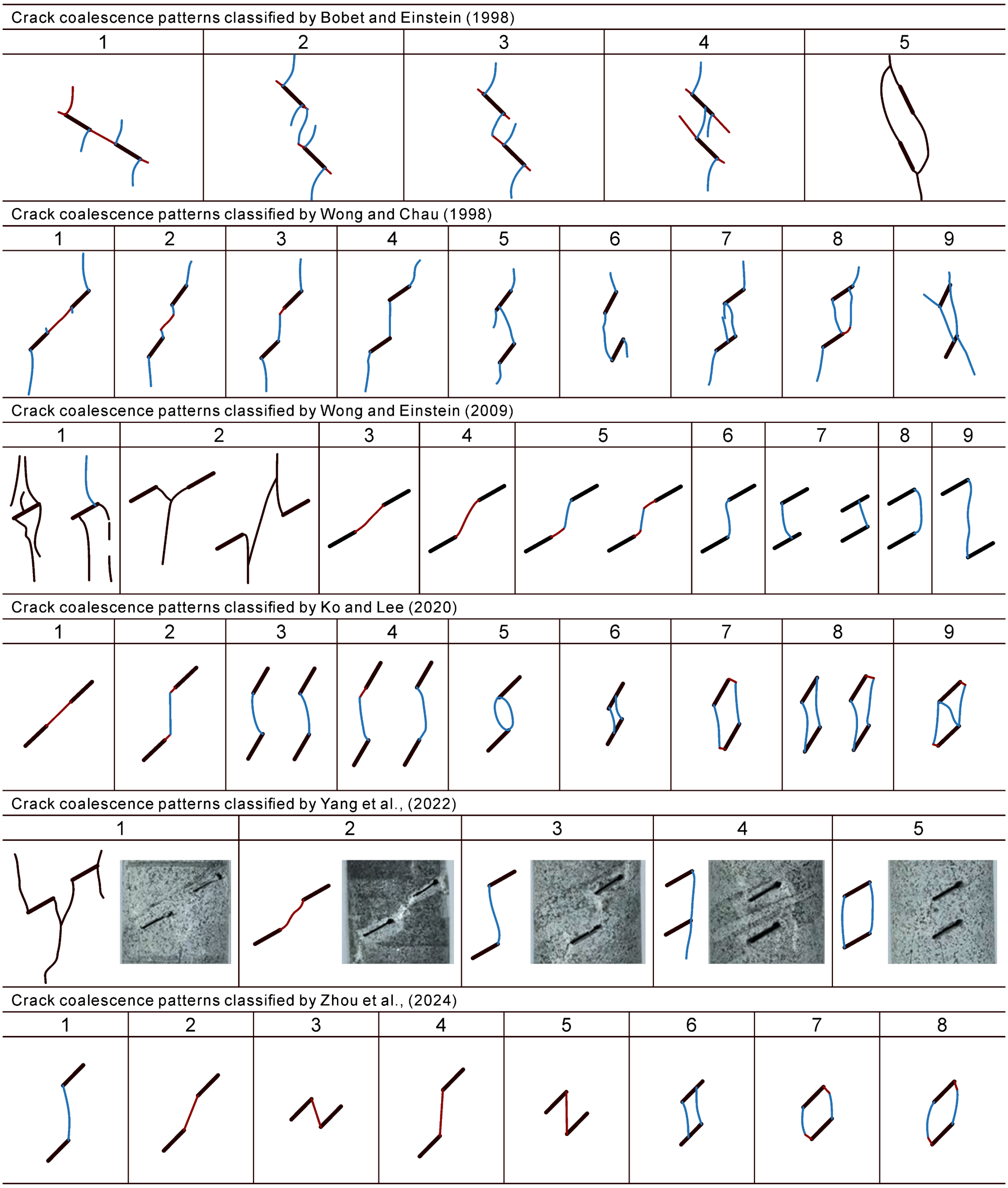

**Fig. 16. Various crack coalescence modes in rock bridges subjected to compression.**

### 7.3. Stress-orientation dependence

Anisotropy constitutes a fundamental attribute of rock masses. Hudson and Harrison (2000) defined anisotropy as rock presenting different properties in different directions. These properties include deformability modulus, strength, brittleness, permeability, and discontinuity frequency. Anisotropy mainly arises from the presence of discontinuities such as joints, bedding, and foliations. Therefore, the discontinuity-defined rock bridge may also present different properties in different directions. Compared to the orientation dependence of discontinuities, investigations into that of rock bridges began much later. Seiki et al. (1993) examined rock mass behaviour using mortar specimens containing eighteen 5 mm-thick urethane-filled slits. These artificial joints were oriented at angles from 15° to 75°, revealing that steeper inclinations significantly enhanced the peak strength and elastic modulus of the material. This study presents an early discovery of the stress-orientation dependence of rock bridges. Another representative early work is from Bagheripour (1996), who documented the stress-orientation dependence of rock bridges. He conducted unconfined compressive tests on rock-like materials containing rock bridges with different joint angles. Results indicated that the failure morphology and strength varied with joint orientation (Fig.17). He tried Jeager's single plane of weakness theory to interpret this but failed, as the geometric descriptors of rock bridges were not controlled.

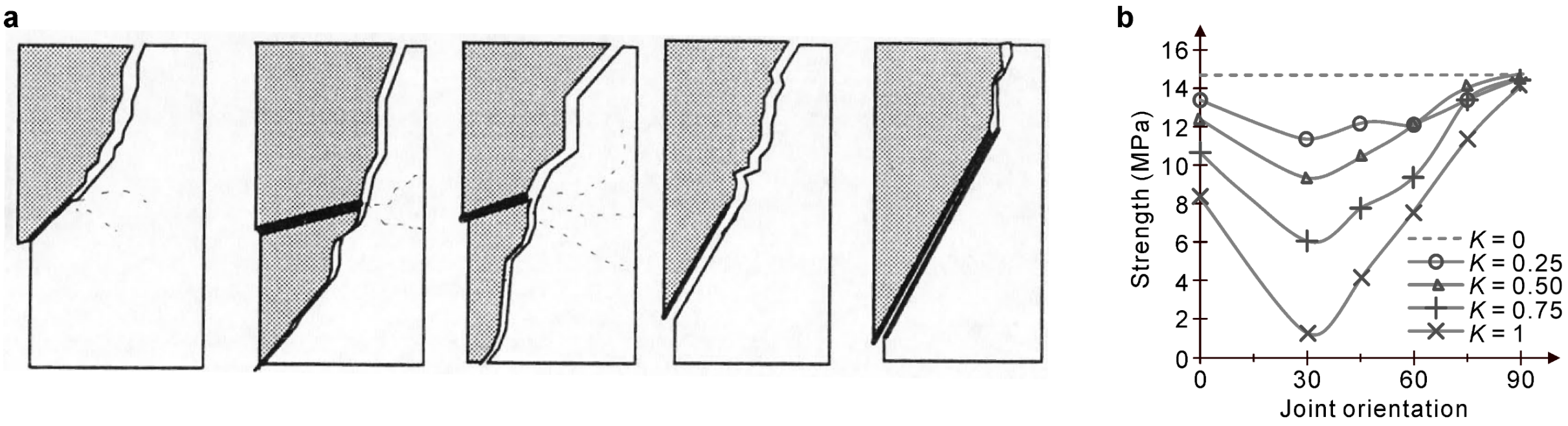

**Fig. 17. Stress-orientation dependence of rock bridges. (a)** Experimental failure morphology. **(b)** Strength varies with joint orientation, modified from Bagheripour (1996).

Subsequent analyses of the stress-orientation dependence of rock bridges rely on observational interpretation. Mughieda and Alzo'ubi (2004) conducted unconfined compressive tests on rock-like materials containing rock bridges with an identical inclination angle of 45° but different bridging angles. They employed the classification by Wong and Chau (1998), and found that a small bridging angle favoured coplanar shear cracks, an intermediate bridging angle favoured tensile and shear cracks, and a large bridging angle favoured tensile cracks. Since then, the effect of bridging angle has been widely studied. For example, Prudencio and Jan (2007) conducted biaxial compressive tests on rock-like materials containing impersistent discontinuities with different orientations. They observed three basic failure modes: failure through a planar surface, stepped failure, and failure by rotation of new blocks, which heavily depends on the bridging angle. Yang et al. (2021) conducted unconfined compressive tests on sandstone rock bridges with different bridging angles. Results indicate that, as the bridging angle increases, the total count of microcracks, along with the proportion of shear and mixed-mode fractures, rises steadily. This drives a transition from simple tensile failure to complex crack coalescence. At the microscopic scale, the bridging angle dictates the stress orientation of individual particles, which subsequently governs the trajectory of macroscopic crack propagation. A recent study on the effect of bridging angle is from Diao et al. (2025), who conducted unconfined compressive tests and simulations on sandstone rock bridges with identical length but different bridging angles. They observed three primary coalescence modes: joint-tip-to-extremity, inter-tip bridge failure, and oblique linear fracturing. Localised wear surfaces are concentrated within the locked mass, the bridge region, or around joint tips. Increasing the bridging angle typically reduces peak stress and promotes concentrated fracturing, while shifting the central displacement field from a shell to a core configuration. Although stress concentrations at bridge ends become more pronounced under loading, their intensity remains constrained by the specific inclination. Finally, normal contact force distribution widens as the peak tangential force aligns more closely with the axial orientation.

In the absence of a rock-bridge mechanics framework, assessing the stress-orientation dependence of rock-bridge behaviour remains problematic, as current analyses rely primarily on observational and taxonomic interpretations rather than mechanistic principles.

### 7.4. Impasse in fundamental research

Lab-scale direct shear or compressive tests on rock bridges typically utilise prefabricated discontinuities; however, the infinite permutations of discontinuity arrangements and boundary conditions yield a vast array of experimental outcomes. Rock-bridge failure mechanisms can be tensile, shear, or their combination; rock-bridge failure morphologies can be a single penetration, an intricate path circumventing it, or an avoidance of it; and measured strengths in shear, compression, or their combination are highly idiosyncratic and sensitive to minor perturbations of experimental scenarios (Sjöberg, 1999; Goodman and Kieffer, 2000; Frayssines and Hantz, 2006; Huang et al., 2015; Afolagboye et al., 2017; Aliabadian et al., 2021). In the absence of a rock-bridge mechanics framework, most studies are confined to phenomenological categorisation and restricted statistical analysis, precluding the effective upscaling of these results to complex natural fracture networks.

## 8. Rock-bridge mechanics: Fundamentals and future

### 8.1. Rock-bridge failure criteria

Bu et al. (2026b) proposed rock-bridge failure criteria. As exemplified by the rock-bridge specimen containing open discontinuities subjected to unconfined compression ($\sigma_1 > \sigma_3 = 0$) in Fig. 18b, following the geometric descriptors of rock bridges proposed by Einstein et al. (1983) in Section 5.1, the ligament forms an angle of $\theta$ to $\sigma_3$, where $\sigma_1$ and $\sigma_3$ denote the maximum and minimum principal stresses, respectively. The dissipation of driving stress by open discontinuities renders the intact rock enclosing the ligament (white rectangle in Fig. 18b) as nearly the sole compression-bearing structure. Focusing on this intact rock, the stress components on the ligament can be expressed by:

$$\begin{cases} \sigma_n = \dfrac{F\cos\theta}{bL} \\ \tau = \dfrac{F\sin\theta}{bL} \end{cases} \tag{50}$$

where $\sigma_n$ is the normal stress, $\tau$ is the shear stress, $F$ is the external force, $L$ is the ligament length, and $b$ is the specimen thickness. As $F$ linearly increases, both $\sigma_n$ and $\tau$ present linear growth patterns following $\tau = \sigma_n\tan\theta$, that is, the stress path on the ligament.

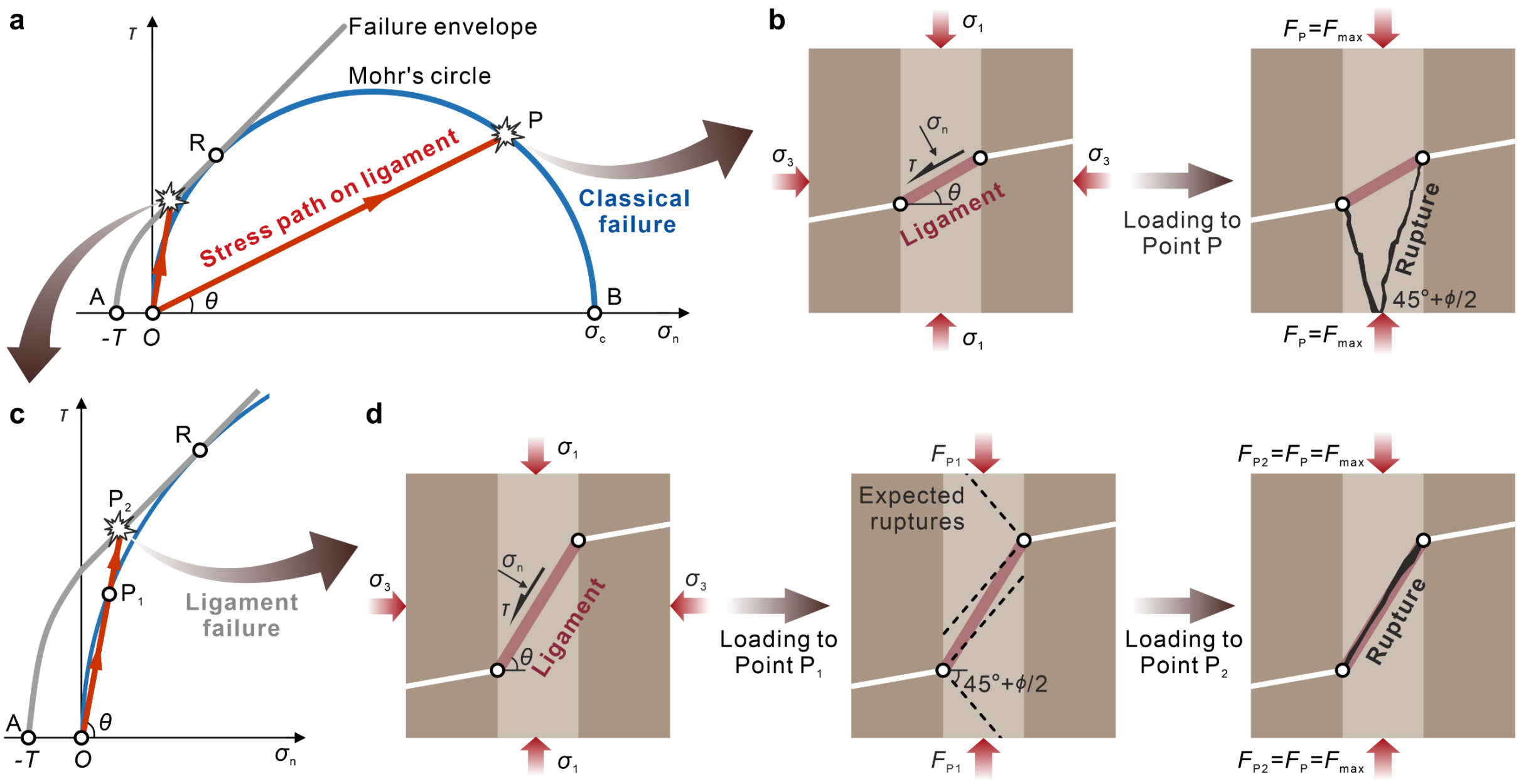

**Fig. 18. Rock-bridge failure criteria. (a)** Stress path on ligament intersecting curve RB at point P corresponds to classical failure. **(b)** Classical failure planes at 45° + $\phi$/2 to $\sigma_3$. **(c)** Stress path on ligament intersecting curve RA at point $P_2$ corresponds to ligament failure. **(d)** Ligament failure follows the ligament. $\tau$, shear stress; $\sigma_n$, normal stress; $\sigma_c$, unconfined compressive strength; $T$, uniaxial tensile strength; $\sigma_1$, maximum principal stress; $\sigma_3$, minimum principal stress, which is 0 for simplicity; $\phi$, internal friction angle; $\theta$, ligament inclination angle; $F$, external force; $F_P$, rock-bridge resistance. Modified from Bu et al. (2026a).

When $\theta$ < 45° + $\phi$/2, the stress path on the ligament increases and intersects the Mohr's circle at point P (Fig. 18a). Isolating the white rectangular rock in Fig. 18b, when the Mohr's circle touches the failure envelope, this intact rock fails along classical rupture planes aligned relative to $\sigma_3$ at 45° + $\phi$/2, since the stress components on the rupture planes, denoted by point R (the tangency point between the Mohr's circle and the failure envelope in Fig. 18a), satisfy the failure criterion. At this moment, the stress components on the ligament are denoted by point

P. Therefore, the intersection point P also indicates the occurrence of classical failure. The coordinates of the intersection point P are the simultaneous solution of the stress path on the ligament and Mohr's circle:

$$\begin{cases} \tau = \sigma_{\mathrm{n}} \tan\theta \\ (\sigma_{\mathrm{n}} - \dfrac{\sigma_{\mathrm{c}}}{2})^2 + \tau^2 = (\dfrac{\sigma_{\mathrm{c}}}{2})^2 \end{cases} \tag{51}$$

where $\sigma_{\mathrm{c}}$ is the unconfined compressive strength. Substituting the derived coordinates of point P into Eq. (50), the corresponding $F_{\mathrm{P}}$ can be obtained, that is, rock-bridge resistance, which equals the maximum force sustained by the specimen $F_{\max}$.

When $\theta \geq 45° + \phi/2$, the stress path on the ligament first intersects the Mohr's circle at point $P_1$ (Fig. 18c); however, the stress components on the plane inclined at $45° + \phi/2$ to $\sigma_3$ are averaged out by the regions outside the intact rock enclosing the ligament, thus no longer reaching the limiting state, and the expected ruptures fail to coalesce with each other (Fig. 18d), resulting in the first intersection $P_1$ no longer signifying failure. With a further increasing $F$, the stress path will intersect the failure envelope at point $P_2$. This second intersection signifies the limiting stress state on the ligament itself, and failure will occur along the ligament rather than along the classical rupture planes. $P_2$ coordinates are calculated by simultaneously solving the stress path on the ligament and the failure envelope:

$$\begin{cases} \tau = \sigma_{\mathrm{n}} \tan\theta \\ \tau = f(\sigma_{\mathrm{n}}) \end{cases} \tag{52}$$

The failure envelope $\tau = f(\sigma_{\mathrm{n}})$ has numerous mathematical expressions, for example, Mohr–Coulomb failure criterion, Hoek–Brown failure criterion, parabolic function, hyperbola function, etc. Substituting the derived coordinates of point $P_2$ into Eq. (50), the rock-bridge resistance $F_{\mathrm{P}}$ can be obtained.

By combining these two mechanisms, rock-bridge failure criteria are established, which can yield rock-bridge resistance and failure modes. Stress paths on ligaments intersecting curve RB correspond to classical failure, while those intersecting curve RA correspond to ligament failure (Fig. 18a). Curve RB is the partial Mohr's circle, while curve RA is the partial failure envelope. They are tangent at point R, corresponding to the critical value of $\theta = 45° + \phi/2$. Both curves RB and RA can be determined by basic rock mechanics tests. Generalising from 2D regular discontinuities in unconfined compression to 3D arbitrary discontinuities in unconfined/confined tension/compression, alongside multiple rock bridges and infilled discontinuities, the rock-bridge failure criteria remain compatible. See Bu et al. (2026b) for more generalisations and experimental validations, as well as the failure criteria for overlapping cases that are not detailed here.

### 8.2. Conceptual advance

Section 3 indicates that rock-bridge definition has been challenged by disciplinary specialisation and conceptual extension. By treating rock bridges as purely geological structures defined by geometric descriptors $\theta$ and $L$, the rock-bridge failure criteria offer a unified analytical framework for rock bridges applicable to multiple disciplines. For example, Bu et al. (2026a) applied the rock-bridge failure criteria to analysing the potential Boot Flake rockslide in an exfoliating cliff of El Capitan within Yosemite Valley, California, USA, and the fault hard-linkage between the Southern San Andreas Fault (SSAF) and Imperial Fault (IF). Although boundary conditions and constitutive laws may change depending on the context, the concept of the failure criteria remains compatible, which can eliminate disciplinary silos and promote knowledge exchange. Additionally, by quantifying rock-bridge resistance magnitude, the rock-bridge failure criteria establish a basis for conceptual extension of rock bridges. As exemplified by the conceptual extension of rock bridges to locked segments, previous views typically assume a simple equivalence between the two, whereas in reality not every rock bridge can be identified as a locked segment. The critical criterion for such a mechanical extension is the resistance provided by the rock bridge. The rock-bridge failure criteria fulfil this evaluative need and consequently clarify whether rock bridges can be extended to locked segments.

Section 4 indicates that the field incorporation of rock bridges has long been hindered by the absence of a rock-bridge mechanics framework, a deficiency addressed by the rock-bridge failure criteria. By requiring only parameters of ligaments, stress, and rocks that are obtainable from basic field investigations and rock mechanics tests, the rock-bridge failure criteria enable the analysis of rock bridges by deriving their resistance magnitude and failure modes, which are critical inputs to assess rock mass stability, as evidenced by the applications to the Boot Flake landslide and the SSAF–IF fault hard-linkage. These applications demonstrate that the rock-bridge failure criteria elevate rock bridges from conceptual entities to analysable objects.

Section 5 indicates that the principal challenge in measuring rock bridges lies in proactive ones, attributed to the difficulties in geometric and mechanical measurements. By characterising rock-bridge profiles as ligament lines in 2D and ligament surfaces in 3D, the rock-bridge failure criteria avoid the conceptual definitions of $RB_{32}$ and $RB_{33}$ in the $RB_{ij}$ system (Tuckey et al., 2012; Elmo et al., 2018), and the statistical uncertainties of discontinuity tip counting in rock mass classification systems (Laubscher, 1990; Kemeny et al., 2003; Elmo, 2023; Xu et al., 2026), while also overcoming the constraints of employing deformation or stress gradients to define rock-bridge profiles (Tang et al., 2024; Xi et al., 2025). While incorporating simplifications, it stems from the rock-bridge definition and employs only measurable parameters. This focus on the rock bridge itself supports more effective subsequent mechanical analysis. Moreover, the indescribable rock bridges prior to failure in fact denote those whose resistance roles in the failure process cannot be determined, namely, mechanical measurements. Elmo (2023) described the contribution of rock bridges to rock mass strength as a potential. The rock-bridge failure criteria enable the quantification of this potential by deriving the resistance magnitude of each rock bridge.

Section 6 indicates that, to date, Lajtai's criterion, Jennings's criterion, and fracture mechanics-based methods, alongside their developments, remain the dominant paradigms in analysing rock bridges. However, their fundamental deficiencies persist. Lajtai's criterion and Jennings's criterion are constrained by the concept of $K$ and their limitations to specific failure mechanisms. Fracture mechanics has difficulty interpreting how a rock bridge resists rupture as a whole, and the solutions are always limited to specific discontinuity arrangements and sensitive to burdensome parameters. In contrast, the rock-bridge failure criteria break free from the framework of $K$, transcend specific failure mechanisms, and apply universally to tensile, shear and mixed-mode failures. By deriving the resistance magnitude and failure modes, the rock-bridge failure criteria elucidate the resistance roles of rock bridges as a whole, addressing the difficulties of fracture mechanics in characterising crack coalescence. Furthermore, the parsimony and low parameter demand enable the rock-bridge failure criteria to be readily deployable in complex and broader applications.

Section 7 indicates that the mechanical behaviour of rock bridges, including resistance magnitude, failure mechanism, fracture morphology, and stress-orientation dependence, is highly complex due to the diversity of rock-bridge configurations and stress-orientation dependence. Focusing on ligaments, the rock-bridge failure criteria establish the relationship between rock-bridge configurations and mechanical behaviour. In this way, following the determination of ligament stress paths, rock-bridge resistance magnitude, failure mechanism (tensile, shear and mixed modes), and fracture morphology (classical, ligament, and cantilever beam failure) can be derived. Specifically, as shown in Fig. 18a, stress paths on ligaments intersecting curve RB indicate classical failure, where rupture planes form an angle of $45° + \phi /2$ to $\sigma_3$; stress paths on ligaments intersecting curve RA indicate ligament failure, where rupture planes align with the ligament; overlapping parts are simplified as cantilever beams, where rupture planes are vertical to discontinuities. Substituting the coordinates of the intersection point between stress paths on ligaments and curve ARB into Eq. (50) yields the rock-bridge resistance. Employing the rock-bridge failure criteria, the stress-orientation dependence can also be clarified. In the recent research by Bu et al. (2026a), as the rotation of stress directions is equivalent to that of rock bridges, stress directions and geometric descriptors were fixed, while rock bridges were rotated. They found that stress direction exerts a profound influence on both the resistance magnitude and failure modes of rock bridges, depending on the ligament inclination angle $\theta$.

### 8.2. Future work

The rock-bridge failure criteria establish a fundamental framework for rock-bridge mechanics by enabling the mechanical analysis of rock bridges. The derived rock-bridge resistance and failure modes have been validated across various conditions including boundary conditions (unconfined/confined compression, tension, and shear), lithologies (rock-like material, sandstone, limestone, granite), and discontinuity arrangements (varied bridging angle, ligament length, infilling materials) (Bu et al., 2026b). Theoretical applications are exemplified through predicting a rockslide and fault hard-linkage (Bu et al., 2026a). The derived analytical solutions for predicting rock-bridge failure are essential for understanding and managing rock mass instabilities. Below are several topics that might be taken up for future research:

(1) Ligament measurement. Einstein et al. (1983) defined a ligament as the line connecting adjacent discontinuity tips. As the density of discontinuities increases, this line will present uncertainty, as equidistant pairings may occur and non-adjacent discontinuity tip connections can be critical to stability. One potential methodology involves constructing an exhaustive set of ligaments from all tip combinations for systematic evaluation. This uncertainty is exacerbated in 3D, where ligament surfaces can be defined through various combinations of tips and lines, necessitating a more rigorous characterisation of ligament morphology.

(2) Stress path determination. While numerous methods exist for measuring *in situ* stress, the precise characterisation of stress paths on ligaments remains absent. Section 8.1 assumes linearly increasing stress paths on ligaments for illustration. In applications of the theory, such as field-scale rock bridges and rock masses containing multiple rock bridges, more robust methods to obtain the stress paths on ligaments should be adopted. Alternative methods include: (1) Analytical method. Küntz et al. (1997; 1998) proposed analytical methods for calculating stress fields in cracked media by establishing the correspondence between equilibrium equations for an incompressible elastic solid and a viscous fluid in steady-state, from which stress paths on ligaments can be obtained. (2) Experimental method. DIC enables the extraction of strain fields that can be converted into stress fields using Hooke’s law, which further yields stress paths on ligaments. Bu et al. (2026a) used this method to obtain the experimental stress paths on ligaments, which is also compatible with multiple rock bridges. (3) Numerical method. Numerical modelling of rock masses allows the extraction of stress fields (Bu et al., 2025; 2026c; 2026d); subsequently, the stress paths on ligaments can be obtained. This method was used in both the Boot Flake rockslide case and the fault linkage case in Bu et al. (2026a). Future research should focus on developing specialised *in situ* measurement techniques dedicated to rock bridges, as these would prioritise specific data points on ligaments rather than requiring exhaustive regional surveys.

(3) Microcracking on ligaments. Classical failure and ligament failure correspond to distinct macroscopic failure modes. Although these modes have been identified across compression, tension, and shear tests, their underlying microcracking processes remain poorly understood, especially in rock bridges. X-ray computed tomography (CT) offers a viable diagnostic tool and has been utilised in studies of rock-bridge failure (Yang et al., 2018b; Prastyani et al., 2025); however, existing research is fragmented and restricted to limited rock-bridge settings. Integrating CT with the proposed rock-bridge failure criteria may provide a pathway to elucidate the microscopic origins of classical failure and ligament failure modes, thereby offering deeper insights into rock-bridge failure.

## 9. Conclusions

Based on a manual review of each of the 2538 rock-bridge research spanning 1962–2025, we conclude that:

(1) Research on rock bridges has a long history and can be broadly divided into four developmental phases: conceptual genesis and theoretical grounding (1962–1970), practical implementation (1970–1990), methodological maturation (1990–2018), and expansion alongside debate (2018–2025).

(2) Rock bridges are defined geometrically but have undergone decades of disciplinary specialisation and mechanical extension. The specialisation has resulted in disciplinary silos, while the lack of a rock-bridge mechanics framework has led to misunderstanding of the term and its extended meanings.

(3) Rock bridges are important in geological processes (landslides, earthquakes, tectonics, geomorphology, hydrocarbon reservoirs, etc.) and rock engineering applications (mining excavation, slope stability, dam foundations, etc.). Their mechanical behaviours are highly complex, modulated by multiple factors (geological structures, stress states, lithology, topography, hydrogeology, etc.), yet a unified mechanical paradigm remains elusive.

(4) Despite extensive forensic and some proactive measurement of rock bridges, alongside probabilistic and statistical methods, measuring mechanically significant rock bridges remains challenging, primarily due to the elusive relationship between rock-bridge configurations and mechanical behaviour.

(5) Lajtai's criterion, Jennings's criterion, fracture mechanics frameworks, and their developments remain the mainstream theories in the rock-bridge domain. However, their limitations to specific failure modes and reliance on the concept of discontinuity persistence prevent them from describing complex mechanical behaviour of rock bridges, leaving the domain without a feasible mechanical framework.

(6) Even at a controllable laboratory scale, artificially notched rocks with various discontinuity arragements under shear and compression present puzzling strengths and intricate fracturing morphologies. Given the absence of a rock-bridge mechanics framework, prevailing studies remain largely restricted to phenomenological classification and descriptive statistics, hindering the integration of that lab-scale work into field-scale natural fracture networks.

Rock-bridge research faces these impasses that all highlight the need for a mechanical framework to analyse rock bridges. Recent advances in rock-bridge failure criteria promise to serve as this framework to address these impasses, as they established the relationship between rock-bridge configurations and mechanical behaviour. The derived rock-bridge resistance magnitude and failure modes are essential inputs to assess rock mass stability, thereby elevating rock bridges from conceptual entities to analysable objects. However, ligament measurement, stress path determination, and microcracking on ligaments warrant further study.

## Acknowledgements

This study was financially supported by the National Natural Science Foundation of China grants 41977249 and 42090052, the Swiss National Science Foundation Postdoc.Mobility fellowship P500-2_241983, and the China Scholarship Council grant 202204910040. Authors extend their appreciation to H.H. Einstein at MIT, B.D. Collins at USGS, V. Merrien-Soukatchoff at CNAM, and M. Grenon at ULaval for the discussions and comments.

## Author contributions

F. Bu conceived the idea, analysed the theoretical, experimental, numerical, and field-based data, and wrote the manuscript. M. Jaboyedoff supervised the project, discussed the results, and finalised the manuscript. L. Xue discussed the results. All the authors contributed to the discussion and comments on the manuscript.

## Competing interests statement

The authors declare no competing interests.

**Correspondence and requests for materials** should be addressed to Fengchang Bu.